%% file: QCD-10-037_temp.tex
\pdfoutput=1

\documentclass[11pt,twoside,a4paper,cmspaper,final,collab]{cms-tdr}
\def\svnVersion{74104:74142}\def\cmsCernNo{2011-128}\def\cmsCernDate{\today}\def\cmsMessage{Submitted to Physical Review D}

\begin{document}\cmsNoteHeader{QCD-10-037}

\hyphenation{had-ron-i-za-tion}
\hyphenation{cal-or-i-me-ter}
\hyphenation{de-vices}
\RCS$Revision: 74142 $
\RCS$HeadURL: svn+ssh://alverson@svn.cern.ch/reps/tdr2/papers/QCD-10-037/trunk/QCD-10-037.tex $
\RCS$Id: QCD-10-037.tex 74142 2011-08-09 20:35:54Z alverson $
\providecommand{\jetphox}{\textsc{Jetphox}\xspace}
\providecommand{\etgamma}{\ensuremath{E_\mathrm{T}}}
\providecommand{\xtgamma}{\ensuremath{x_\mathrm{T}}}
\providecommand{\Iso}{\ensuremath{\mathrm{ISO}}}
\providecommand{\etagamma}{\ensuremath{\eta}}
\providecommand{\ptgamma}{\ensuremath{p_\mathrm{T}}}
\providecommand{\ecal}{\textsc{ECAL}\xspace}
\providecommand{\hcaliso}{\ensuremath{\mathrm{Iso}_\mathrm{HCAL}}}
\providecommand{\trkiso}{\ensuremath{\mathrm{Iso}_\mathrm{TRK}}}
\providecommand{\ecaliso}{\ensuremath{{\rm Iso}_\mathrm{ECAL}}}
\providecommand{\HCAL} {\textsc{HCAL}\xspace}
\providecommand{\APD}{\textsc{APD}s\xspace}
\newlength\figwid
\ifthenelse{\boolean{cms@external}}{\setlength\figwid{0.46\textwidth}}{\setlength\figwid{0.4\textwidth}}
\newlength\onecolfigwid
\ifthenelse{\boolean{cms@external}}{\setlength\onecolfigwid{0.95\columnwidth}}{\setlength\onecolfigwid{0.4\textwidth}}
\title{Measurement of the Differential Cross Section for Isolated Prompt Photon Production in $\Pp\Pp$ Collisions at 7~TeV}

\date{\today}

\abstract{
A measurement of the differential cross section for the inclusive production
of isolated prompt photons in proton-proton collisions at a centre-of-mass
energy of 7~TeV is presented.
The data sample corresponds to an integrated luminosity of 36~pb$^{-1}$
recorded by the CMS detector at the LHC.
The measurement covers the pseudorapidity range $|\eta|<2.5$ and the
transverse energy range $25 < E_\mathrm{T} < 400$~GeV, corresponding to the
kinematic region $0.007 < x_\mathrm{T} < 0.114$.
Photon candidates are identified with two complementary methods, one based on
photon conversions in the silicon tracker and the other on isolated energy
deposits in the electromagnetic calorimeter.
The measured cross section is presented as a function of $E_\mathrm{T}$
in four pseudorapidity regions. The next-to-leading-order
perturbative QCD calculations are consistent with the measured cross section.
}

\hypersetup{%
pdfauthor={CMS Collaboration},%
pdftitle={Measurement of the Differential Cross Section for
Isolated Prompt Photon Production in pp Collisions at 7~TeV},%
pdfsubject={CMS},%
pdfkeywords={CMS, QCD, photon, conversion, isolation sum, endcap}}

\maketitle 

\section{Introduction \label{sec:intro}}
The measurement of isolated prompt photon production in proton-proton
collisions provides a test of perturbative quantum chromodynamics
(pQCD)~\cite{owens0,owens,jetphox,jetphox2}.
The cross section measured in $\Pp\Pp$ collisions also serves as a reference
for similar measurements in heavy ion collision data~\cite{Phenix}.
In addition, isolated prompt photon
production represents a background to searches for new phenomena involving
photons in the final state, including Higgs boson production~\cite{PTDR2}.
At the Large Hadron Collider (LHC)~\cite{LHCmachine}, a significant increase of
centre-of-mass energy with respect to previous collider
experiments~\cite{RHIC,ISRphoton,1984photonreview,SppSphoton,d0photon,cdfphoton}
allows for the exploration of new kinematic regions in the hard scattering
processes in hadron-hadron collisions. In high-energy $\Pp\Pp$
collisions, single prompt photons are produced directly in $\cPq\cPg$ Compton
scattering and $\cPq\cPaq$ annihilation, and in the fragmentation of partons
with large transverse momentum. Photons are also produced in the decay of
hadrons, mainly $\Pgpz$ and $\Pgh$ mesons, which can mimic prompt production.
This background contamination can be estimated from data using photon
identification characteristics, such as
electromagnetic shower profile, extra energy surrounding the photon candidate
(called ``isolation sum'' in this article), or kinematic variables of converted
photons.

Both the CMS and ATLAS Collaborations have performed measurements of the
differential cross section of isolated prompt photon production with
data collected in 2010~\cite{ref:QCD-10-019,ref:ATLASInclPho,ref:ATLASnew}.
The CMS Collaboration reported a measurement for
photons with $21<\etgamma<300\GeV$ and $|\etagamma|<1.45$ with an integrated
luminosity of 2.9\pbinv and exploited the electromagnetic shower profile
to estimate the background contribution~\cite{ref:QCD-10-019}.
Here, $\etgamma=E\,\sin\theta$ and $\etagamma=-\ln[\tan(\theta/2)]$,
where $E$ is the photon energy and $\theta$ is the polar angle of the
photon momentum measured with respect to the counterclockwise beam direction.
The measurement reported in this paper extends the previous CMS measurement to
wider ranges of transverse energy ($\etgamma$ = 25--400\GeV) and
pseudorapidity ($\left|\etagamma\right| < 2.5$), corresponding to the
kinematic region $0.007 < \xtgamma < 0.114$, where
$\xtgamma=2\etgamma/\sqrt{s}$.

The background contribution to isolated photons is estimated with two
methods. The ``photon conversion method'' uses the variable $\et/\pt$, the
ratio of the transverse energy measured in the electromagnetic
calorimeter to the transverse momentum measured in the tracker for converted
photons. The ``isolation method'' uses the variable $\Iso$, the isolation sum
measured in the tracker and the electromagnetic and hadronic calorimeters.
The weighted average of the differential cross sections measured with the
two methods is reported as a
function of $\etgamma$ in four intervals of pseudorapidity: $|\etagamma|<0.9$,
$0.9 < |\etagamma|< 1.44$, $1.57 < |\etagamma|< 2.1$, and
$2.1 < |\etagamma|< 2.5$. The size of the converted-photon sample is limited
due to the probability for a photon to convert before reaching the CMS
electromagnetic
calorimeter and the relatively small conversion reconstruction efficiency.
On the other hand, the signal purity obtained with the photon conversion
method is very high at low photon $\etgamma$, while the isolation method
is less effective at separating signal from background at low photon
$\etgamma$. A combination of the cross section measurements minimises
statistical and systematic uncertainties and yields better overall performance.

This paper is organised as follows.
Section~\ref{sec:cms} describes the relevant CMS detector components.
Sections~\ref{sec:sample},~\ref{sec:evtsel}, and \ref{sec:phoid} list the data
and simulation samples, the event selections, and the photon identification
criteria that are applied in the analysis, respectively.
Sections~\ref{sec:yield} and \ref{sec:eff} detail the methods used to
extract the signal photon yield and the estimation of signal
efficiency.
Section~\ref{sec:sys} describes the sources of systematic uncertainties on
the cross section measurement and
Section~\ref{sec:results} presents the measured differential cross section.
Section~\ref{sec:dataToTheory} discusses the
comparison of experimental measurements with next-to-leading-order (NLO) pQCD
calculations.

\section{The CMS Detector}
\label{sec:cms}
The CMS detector is a general purpose detector built to explore physics at
the \!\TeV scale and is described in detail in Ref.~\cite{cmsjinst}.
A brief description of the main components that are relevant to the present
analysis is provided here.

The electromagnetic calorimeter (\ecal) consists of nearly $76\,000$
lead tungstate crystals that provide coverage in pseudorapidity
$|\eta| < 1.479$ in a cylindrical barrel region and $1.479 < |\eta| < 3.0$ in
two endcap regions. The crystals are 25.8 $X_{\rm 0}$ long in the barrel and
24.7 $X_{\rm 0}$ long in the endcaps, where $X_{\rm 0}$ denotes the radiation
length.
In the barrel region, the transverse distance from the interaction point to
the front face of crystals, with a size of $ 22 \times 22\mm^2$, is 1.29~m,
corresponding to a granularity of $\Delta\eta\times\Delta\phi=0.0174\times 0.0174$.
In the endcap region, the front face of the crystals is
$ 28.62 \times 28.62\mm^2$ and the distance from the interaction point to
the front face is 3.15~m.
Throughout this paper, $\phi$ is the azimuthal angle measured in radians in
the plane transverse to the beam, from the direction pointing to the centre of
the LHC ring toward the upward direction.
A preshower detector consisting of two planes of
silicon strip sensors that are interleaved with a total of 3~$X_{\rm 0}$ of lead
(2 $X_{\rm 0}$ in the front and 1 $X_{\rm 0}$ after the first silicon plane)
is located in front of the \ecal endcaps, covering $1.653 <|\etagamma|< 2.6$.
Avalanche photodiodes (\APD) are used to detect the scintillation light in
the barrel region, while vacuum phototriodes are used in the endcap
region.
The \ecal has an ultimate energy resolution better than 0.5\% for unconverted
photons with $\etgamma$ above $100\GeV$~\cite{ref:ECAL-en-res}.
In 2010 collision data, for $\et>20 \GeV$, this resolution is already better
than 1\% in the barrel~\cite{ref:EGM-10-003}.

The \ecal is surrounded by a brass/scintillator sampling hadronic calorimeter
(\HCAL) with a coverage up to $|\eta|<3$. The scintillation light is converted
by wavelength-shifting fibres that are read out with hybrid photodiodes.
The \HCAL is subdivided into towers with a segmentation of
$\Delta \eta \times \Delta \phi = 0.087\times0.087$ at central rapidity
($|\eta| < 1.74$) and $0.09\times0.174$ to $0.35\times0.174$ at forward rapidity
($1.74 < |\eta| < 3$).

A silicon tracker is located inside the \ecal.
The tracker consists of two main detectors: three barrel layers and two
endcap disks per side of silicon pixel detectors, covering
the region from $4\cm$ to $15\cm$ in radius, and within $49\cm$ on either
side of the nominal collision point along the LHC beam axis; ten barrel layers
and twelve endcap disks per side of silicon strip detectors, covering the
region from $25$ to $110\cm$ in radius, and within $280\cm$ on either
side of the nominal collision point along the LHC beam axis. The tracker
acceptance extends up to a pseudorapidity of $\left | \eta \right | = 2.5$.
 The tracker, \ecal, and \HCAL are immersed in a 3.8~T axial magnetic field,
which enables the measurement of charged particle momenta over more
than four orders of magnitude, from less than $100\MeV$ to more than
$1\TeV$, by reconstructing their trajectories as they traverse the inner
tracking system. With the silicon tracker, the transverse momentum resolution
for high-momentum tracks (100\GeV) is around 1--2\% up to
$\left | \eta \right |$ = 1.6; beyond this $\eta$ value it degrades due to the
reduced lever arm.

\section{Data and Simulation Samples \label{sec:sample}}

The data sample used in this analysis corresponds to a total integrated
luminosity of $(35.9\pm 1.4)\pbinv$~\cite{lumi} recorded in
2010 with the CMS detector.
The simulated samples were generated with \PYTHIA
version 6.4.20~\cite{pythia64}, the CTEQ6L~\cite{cteq6} parton
distribution functions (PDFs), and the Z2 parameter set~\cite{pythia-z2}.
Generated events are passed through the full \GEANTfour~\cite{geant4} simulation
of the CMS detector and are then reconstructed using the same algorithm as for
the data.
For the simulation of the signal and background, two sets of samples
generated with \PYTHIA are used: one containing direct photons produced
in $\cPq\cPg$ Compton scattering and $\cPq\cPaq$ annihilation, and a second one
generated with all $2\rightarrow 2$ QCD processes that include photons from
initial- and final-state radiation (ISR and FSR), photons from parton
showers, and photons from neutral meson decays.
Isolated direct photons, ISR and FSR photons, and photons from parton showers
are treated as signal, while all other photons are considered to be
background. In the simulation, a signal photon must have an isolation sum of
less than 5\GeV.
The isolation sum is calculated as the sum of the \et of all
charged and neutral particles, after removing the photon, within a cone of
$R\equiv\sqrt{(\eta-\eta^{\cPgg})^2 + (\phi-\phi^{\cPgg})^2}=0.4$,
$\eta^{\cPgg}$ and $\phi^{\cPgg}$ being the coordinates of the photon.
The 5\GeV threshold at the generator level was chosen to ensure greater than
95\% efficiency for direct photons and minimise dependence of the efficiency
on the variation of underlying event models.

\section{Event Selection \label{sec:evtsel}}

Events with high-\et\ photons are selected online by a two-level trigger
system.
At the first level, the \et sum of two neighbouring \ecal\ trigger
towers, a trigger tower being a $5 \times 5$ crystal matrix,
is required to be above 8\GeV. The events that satisfy this selection are
passed on to the second trigger level, the High Level Trigger (HLT).
In the HLT, the energy measured in the crystals is clustered using the same
clustering algorithm as for the offline photon reconstruction~\cite{ref:EGM-10-005,ref:EGM-10-004}.
The events having at least one reconstructed electromagnetic cluster
with an \et above a programmable threshold ($E_\mathrm{T}^\mathrm{HLT}$)
are accepted. In this analysis,
$E_\mathrm{T}^\mathrm{HLT}$ of 20, 30, 50, or 70\GeV are used, depending on
the run period.
Owing to the increase of the LHC instantaneous luminosity and the limited
available trigger bandwidth, different rate-reduction factors
were applied to the triggers at 20, 30, and 50\GeV.
Only data collected without the application of rate-reduction factors are
used, therefore the data samples for events with photons with
$\etgamma < 80\GeV$ correspond to smaller effective integrated luminosities,
as listed in Table~\ref{tab:lumi}.
Events not coming from $\Pp\Pp$ collisions, such as those from beam-gas
interactions or beam scraping in the transport system near the interaction
point, which produce considerable activity in the pixel detector, are removed
by requiring a good primary interaction vertex to be reconstructed.
Such vertices must have at least three tracks and must be within 24\cm (2\cm)
 of the nominal centre of the detector along (perpendicular to) the beam axis.
The efficiency for reconstructing a primary interaction vertex is greater
than 99.5\%~\cite{ref:TRK-10-005}.
In addition, at least 25\% of the reconstructed tracks in the event are
required to satisfy the quality requirements given in Ref.~\cite{highPurity}.

\begin{table}[htbp]
\caption{Effective integrated luminosity for each photon \etgamma\ range.}
\begin{center}
\begin{tabular}{ l r }
\hline
\hline
 \multicolumn{1}{c}{\etgamma\ (\!\GeV)} & \multicolumn{1}{c}{Integrated luminosity (\!\pbinv)} \\
\hline
   25--35 & $2.4\pm 0.1$ \\
   35--55 & $8.2\pm 0.3$ \\
   55--80 & $17.6\pm 0.7$ \\
   $>80$  & $35.9\pm 1.4$ \\
\hline
\hline
\end{tabular}
\end{center}
\label{tab:lumi}
\end{table}

\section{Photon Reconstruction and Identification \label{sec:phoid}}

Photon showers deposit their energy in several crystals in the \ecal. The
presence of material in front of the calorimeter may result in photon conversions.
 Because of the strong magnetic field, the energy deposited in \ecal by converted
photons can be spread in $\phi$. The energy is therefore clustered at the electromagnetic
calorimeter level by building a cluster that is extended in
$\phi$, thus minimising the cluster containment variations due to
electromagnetic interactions in the tracker material. The threshold for crystals
to be included in the cluster is approximately 1\GeV in transverse energy. In
the barrel region of \ecal, clusters are formed from the energy sum in a rectangular
strip of 5 crystals along $\eta$ and up to 35 crystals in $\phi$.
In the endcap region of \ecal, clusters comprise one or more contiguous arrays of
5$\times$5 crystals. Endcap cluster positions are extrapolated to the preshower
where preshower clusters are built. The total endcap cluster energy is the sum of
cluster energies in the endcap crystals and preshower.

Energy corrections are applied to the clusters to take into account
the interactions with the material in front of \ecal\ and shower containment;
the corrections are parametrised as a function of cluster size, \etgamma, and
\etagamma, and are on average 1\%~\cite{ref:EGM-10-003}. The
corrections include the following steps:
\begin{itemize}
\item A compensation of the $\eta$ dependence
of the lateral energy leakage since the axes of the truncated-pyramid
shaped barrel crystals make an angle of 3$^\circ$ with respect to the
vector from the nominal interaction vertex, in both the $\eta$ and $\phi$
projections. This correction is applied only to barrel clusters.

\item A correction to compensate for interactions with material in front of
\ecal. Since these interactions spread energy mainly in the $\phi$ direction,
this loss can be parametrised as a function of the ratio of the cluster size
in $\phi$ to its size in $\eta$.
\item A residual correction that is a function of the cluster $\etgamma$ and
$\etagamma$, to compensate for variations along $\eta$ in the amount of
material and the dependence on $\etgamma$ of the bremsstrahlung and conversion
processes.
\end{itemize}
A photon candidate is built from the energy-corrected cluster, and the
photon momentum is calculated with respect to the location of the
reconstructed primary interaction vertex. If multiple vertices are
reconstructed, the vertex with the largest scalar sum of the transverse
momenta of the associated tracks ($\Sigma \pt$) is selected.

The timing of the \ecal signals is required to be consistent with that of
collision products~\cite{CRAFT-ECAL-timing}.
Topological selection criteria are used to suppress direct interactions
in the \ecal \APD~\cite{ref:NOTE-2010/012}. The residual contamination
has an effect smaller than 0.2\% on the measured cross section over the entire
$\etgamma$ range considered. Contamination from noncollision backgrounds is
estimated to be negligible~\cite{ref:EGM-10-006}.

Photons are required to have a transverse energy $\etgamma > 25\GeV$ since above
25\GeV the trigger efficiency is approximately 100\% for both the barrel and
the endcap photons (Section~\ref{sec:eff}).
The measurements are performed in four photon pseudorapidity intervals:
$|\etagamma|<0.9$ (central barrel), $0.9 < |\etagamma|< 1.44$ (outer barrel),
$1.57 < |\etagamma|< 2.1$ (low-$\eta$ endcaps), and $2.1 < |\etagamma|< 2.5$
(high-$\eta$ endcaps).
This definition excludes the transition region between the barrel and the
endcaps ($1.44 < |\etagamma| < 1.57$) and the region outside of the tracker
coverage ($|\etagamma| > 2.5$). The central barrel has 1--1.5 $X_{\rm 0}$ less
material in front of the \ecal than the outer barrel,
while the low-$\eta$ endcaps have about 0.5 $X_{\rm 0}$ more material than the
high-$\eta$ endcaps, which motivates the subdivision of the barrel and the endcaps.

As mentioned in Section~\ref{sec:intro}, the major source of background comes
from the decays of hadrons (such as $\Pgpz\rightarrow\cPgg\cPgg$) and
nonisolated photons produced by the fragmentation of quarks or gluons.
The photons from hadron decays tend to produce a wider shower profile since
hadrons are
massive and give a nonzero opening angle between the photon daughters.
In addition, both the decay photons and nonisolated fragmentation photons
are accompanied by a number of neutral and charged hadrons that deposit
energy in the \ecal and \HCAL and leave multiple tracks in the tracking system.
Based on these differences between signal and background, several photon
identification variables are used in this analysis:
\begin{itemize}
\item $H/E$: the ratio of the energy deposited in the \HCAL to the energy
deposited in the \ecal inside a cone of $R<0.15$ centred on the reconstructed
photon direction.

\item \texttt{$\sigma_{\eta\eta}$}: a modified second moment of the
electromagnetic energy cluster about its mean $\eta$ position. This quantity
is computed with logarithmic weights and is defined as
\begin{gather*}
\label{eq:sigmaietaieta}
\sigma^2_{\eta\eta} = \frac{\sum_{i}^{5\times5} w_i(\eta_{i}-\bar{\eta}_{5\times5})^2}{\sum_{i}^{5\times5}w_i},\\
w_i = {\rm max}\left(0, 4.7+\ln \frac{E_i}{E_{5\times5}}\right),
\end{gather*}
where $E_{i}$ and $\eta_{i}$ are the energy and pseudorapidity of the
$i^{\text{th}}$ crystal within a matrix of $5\times5$ crystals centred on the
cluster seed and $E_{5\times5}$ and $\bar{\eta}_{5\times5}$ are the energy sum
of the matrix and the weighted average of the pseudorapidities of the same
group.

\item \texttt{\trkiso}: the sum of the transverse momenta (\pt) of all
tracks in a hollow cone $0.04<R<0.4$ drawn around the photon direction.
The tracks pointing to a rectangular strip of width
$\left|\Delta \eta\right|=0.015$ centred around the photon position are
removed from the sum in order to recover possible photon conversions. In
addition, tracks with a transverse (longitudinal) impact parameter above 0.1
(0.2)\cm are not included.

\item \texttt{\ecaliso}: the sum of the \et in the individual \ecal crystals
located in a hollow cone, with an inner radius of 3.5 crystals and an outer
radius of $R=0.4$, drawn around the \ecal cluster.
The \et deposited in a strip of width $\left|\Delta \eta\right| = 2.5$
crystals centred on the photon position is subtracted from the sum to
exclude possible photon conversions.

\item \texttt{\hcaliso}: the sum of the \et in the  \HCAL towers in a hollow cone
$0.15 < R < 0.4$ centred on the \ecal cluster.

\end{itemize}

The signal photons are expected to have smaller values of $H/E$,
\texttt{$\sigma_{\eta\eta}$}, \texttt{\trkiso}, \texttt{\ecaliso}, and
\texttt{\hcaliso} compared to the background photons.
The selection criteria for the two methods are slightly different and are
described in detail in Section~\ref{sec:yield}. 

\section{Extraction of the Prompt Photon Yield \label{sec:yield}}

The following subsections describe the details of extracting the photon
yield ($N^{\cPgg}$) from the two variables (i) $\etgamma/\ptgamma$,
the ratio of the $\etgamma$ measured in the \ecal to the $\ptgamma$ measured
in the tracker for converted photons, and (ii) $\Iso$, which is
$\trkiso+\ecaliso+\hcaliso$.

The photon conversion method relies on the difference in the shape of the
$\etgamma/\ptgamma$ distributions between the signal and background.
For an isolated prompt photon, the sum of the $\ptgamma$ of the conversion
tracks is on average the same as the energy deposited in the \ecal,
and thus the $\etgamma/\ptgamma$ distribution peaks around one.
For photons produced from the decay of $\Pgpz$ and $\Pgh$ in jets, the
$\ptgamma$ measured from the conversion electron pairs does not account for
the full amount of energy collected in the calorimeter and the
$\etgamma/\ptgamma$ is, on average, above one.

The isolation method relies on the difference in the shape of the $\Iso$
distributions. For a photon signal, only underlying event, pile-up, and
detector noise may contribute to the $\Iso$; the $\Iso$ distribution falls
off quickly at around 5\GeV. For a photon background from neutral hadron
decays, the energy of particles that are produced together with $\Pgpz$ or
$\Pgh$ from the parton fragmentation adds a significant amount of activity
around the decay photon and widens the $\Iso$ distribution.

In each method, one of these variables is chosen as a discriminating
observable. A set of preselection criteria is applied to increase the
signal fraction of the photon sample; the signal-region selection criteria are
listed in Table~\ref{tab:preCuts}. The number of signal events $N^{\cPgg}$
is obtained by fitting the distribution of the discriminating observable as
the sum of two components: signal and background. The shapes of the
 component distributions are taken from simulation and are validated
by methods based on data.

\begin{table*}[htbp]
\caption{Signal-region and sideband-region preselection criteria for the
 photon conversion and isolation methods.}
\begin{center}
\begin{tabular}{ c| r r}
\hline \hline
 Cut &  Signal region & Sideband region\\
\hline
  & \multicolumn{2}{c}{Photon conversion method} \\\cline{2-3}
 $H/E$      & $<0.05$   & $ < 0.05 $  \\
 \trkiso\ (\!\GeV)   &  $ < \left(2.0 + 0.001 \etgamma\right) $
                   & $ \left(2.0 + 0.001 \etgamma\right)$ -- $\left(5.0 + 0.001 \etgamma\right) $ \\
 \ecaliso\ (\!\GeV)   &  $ < \left(4.2 + 0.003 \etgamma\right) $  & $ < \left(4.2 + 0.003 \etgamma\right) $ \\
 \hcaliso\ (\!\GeV)   &  $ < \left(2.2 + 0.001 \etgamma\right) $  & $ < \left(2.2 + 0.001 \etgamma\right) $ \\
 barrel: $\sigma_{\eta\eta} $  & $ < 0.010 $ & 0.010 -- 0.015  \\
 endcap: $\sigma_{\eta\eta} $  & $ < 0.030 $ & 0.030 -- 0.045  \\
\hline
  & \multicolumn{2}{c}{Isolation method} \\\cline{2-3}
 $H/E$      & $<0.05$   & $ < 0.05 $  \\
 barrel: $\sigma_{\eta\eta} $ & $ < 0.010 $ & 0.0110 -- 0.0115 \\
 endcap: $\sigma_{\eta\eta} $ & $ < 0.028 $ & $>$ 0.038       \\
\hline \hline

\end{tabular}
\end{center}
\label{tab:preCuts}
\end{table*}

\subsection{Photon conversion method \label{sec:conv}}
After applying the signal-region preselection criteria in Table~\ref{tab:preCuts},
converted photons are reconstructed by combining the information in the \ecal
and the tracker. The \ecal clusters, built and corrected as described in
Section~\ref{sec:phoid}, are used as starting points for an inward
conversion track search, using the $\etgamma$ of subclusters as an
initial guess for the electron or positron trajectory~\cite{ref:EGM-10-005,ref:NOTE-2006-005}.
The innermost point of the resulting tracks is assumed to be close to the
conversion point and used as seed for outward track search of the other arm
of the conversion.
The pattern recognition includes the average energy loss
for electrons passing through the tracker material.
Once all tracks have been found and the track collection cleaned with
loose selection criteria, pairs of oppositely charged tracks
belonging to the same cluster are considered as possible conversion
candidates. A vertex fit imposing the condition that these tracks be parallel
at the conversion vertex is required to converge with a $\chi^2$ probability
greater than $5\times10^{-4}$. The latter ensures that only good vertices are
retained and random or ill-defined pairs are rejected.
Furthermore, since the method is based on the matching between energy-momentum
of the conversions, the requirement $\etgamma/\ptgamma < 3$ is applied.

In each $\etgamma$ bin, the measured $\etgamma/\ptgamma$ distribution is
fitted using a binned extended maximum likelihood method, with the
likelihood defined as
\[
-\ln{L} = (N_s+N_b) - \sum^n_{i=1} N_i \ln(N_s \mathcal{P}_s^i + N_b \mathcal{P}_s^i),
\label{logLike}
\]
where $N_s$ and $N_b$ are the numbers of expected signal and background
events, $n$ is the number of bins, $N_i$ is the number of observed photon
candidates in the $i^{\text{th}}$ bin, and
$\mathcal{P}_s^i$ and $\mathcal{P}_b^i$ are the signal and background
probability density functions integrated over the $i^{\text{th}}$ bin.

Both the signal shape $\mathcal{P}_s$ and the background shape
$\mathcal{P}_b$ are extracted from simulations.
An alternate background shape is extracted from background-enriched data
that are selected by defining a two-dimensional sideband region in the
$\sigma_{\eta \eta}$-\trkiso\ plane according to the definitions given in
Table~\ref{tab:preCuts}. In this sideband region, photon
candidates satisfy loose but fail tight criteria on $\sigma_{\eta \eta}$
and \trkiso; the other requirements are the same as those for
the signal region.
The comparison between the $\etgamma/\ptgamma$ background distributions
obtained from simulation and data gives a measure of the
discrepancy between simulation and data and is used to quantify the
systematic uncertainty related to the modelling of the background shape.
More details are found in Section~\ref{sec:sysconv}.

\begin{figure*}[htbH]
\centering
\includegraphics[width=0.85\figwid,angle=90]{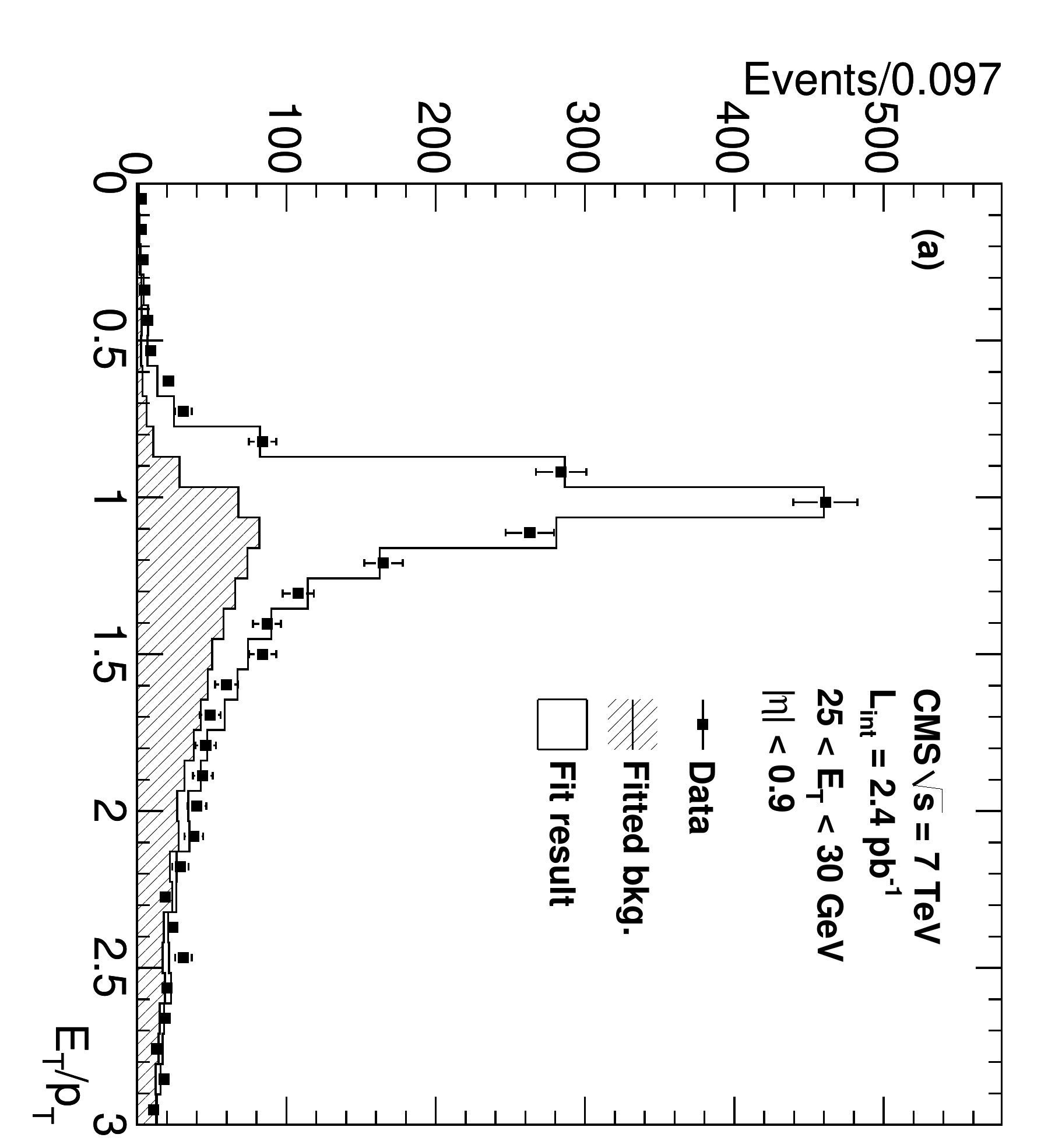}
\includegraphics[width=0.85\figwid,angle=90]{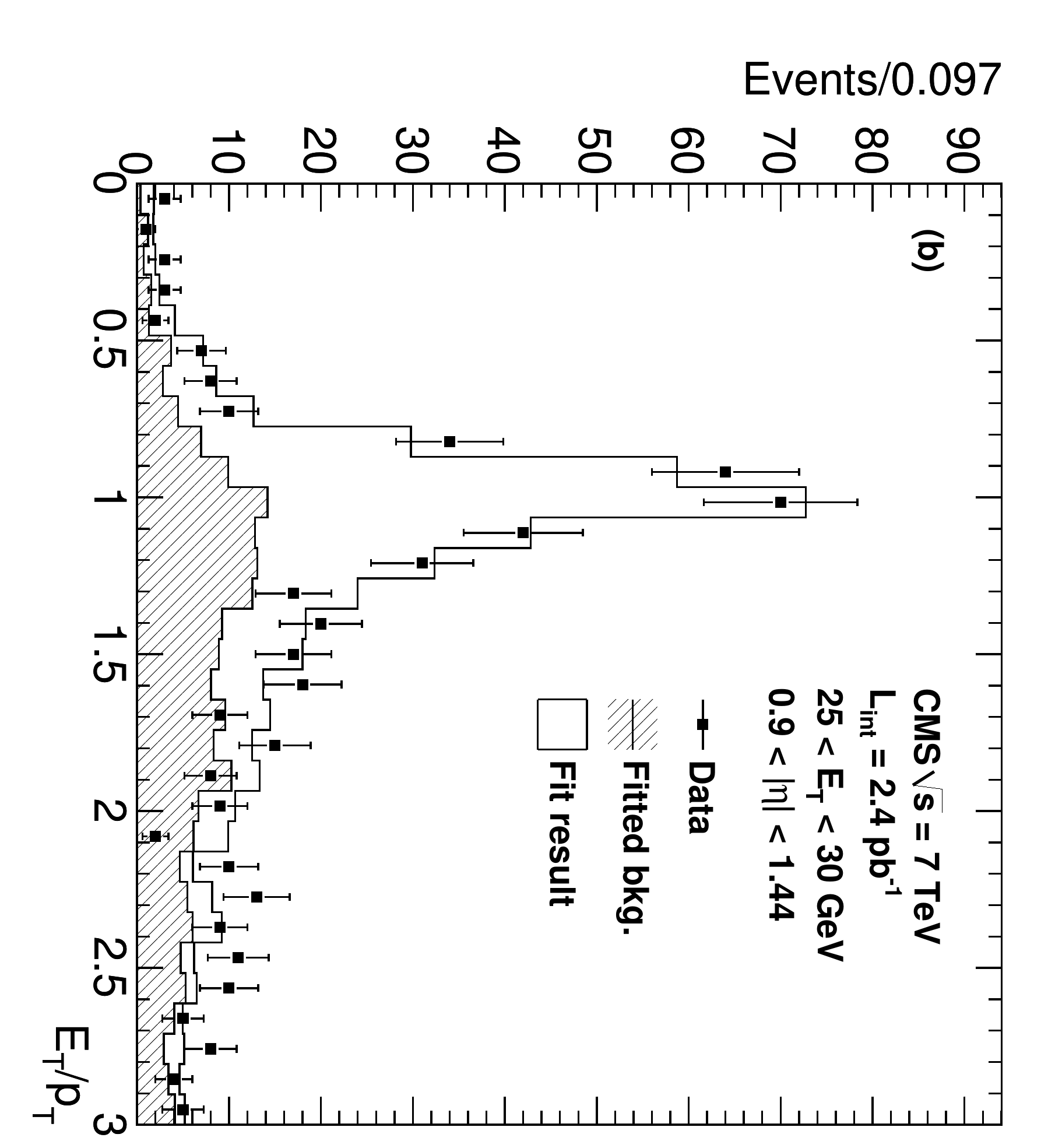}\\
\includegraphics[width=0.85\figwid,angle=90]{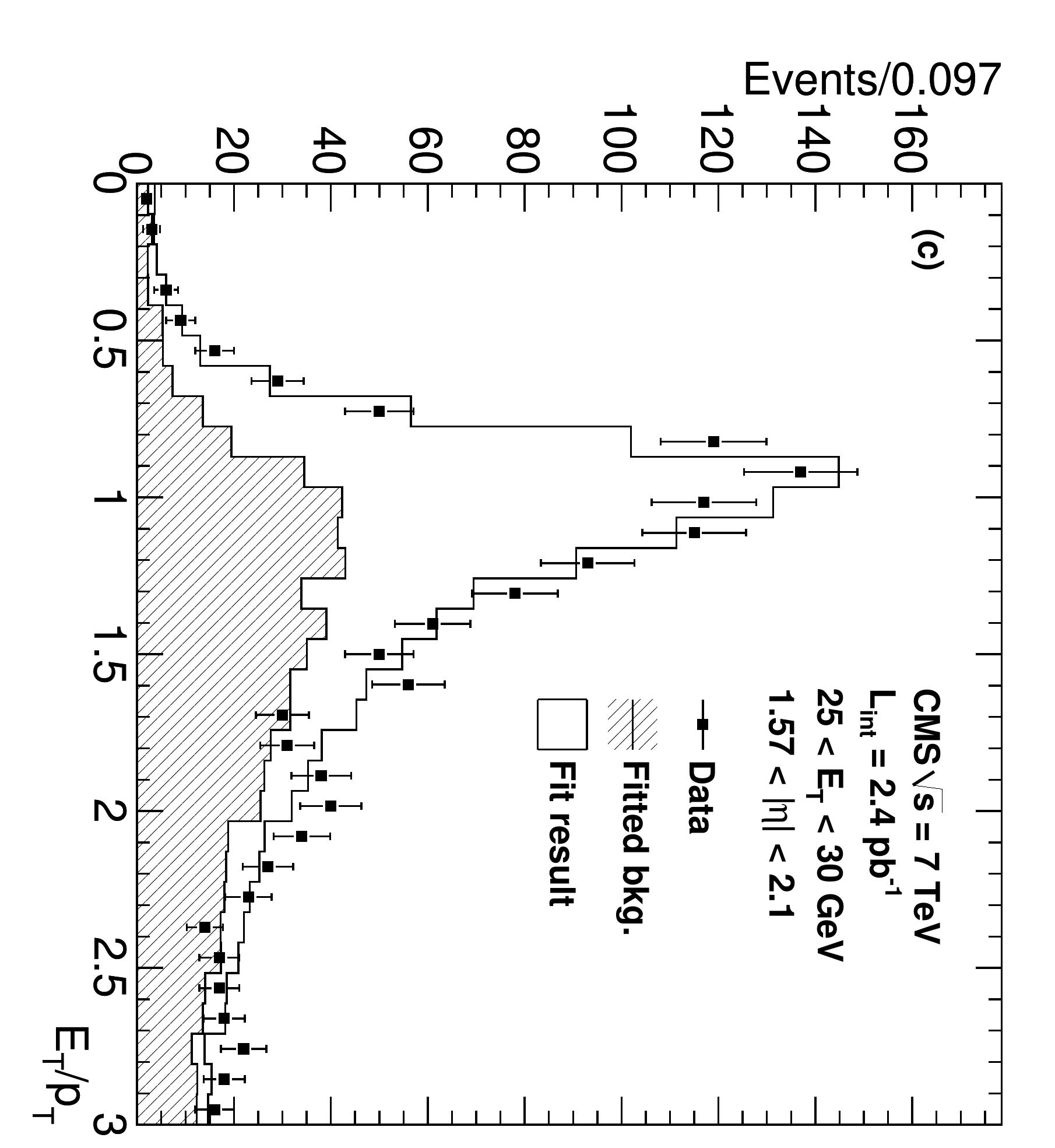}
\includegraphics[width=0.85\figwid,angle=90]{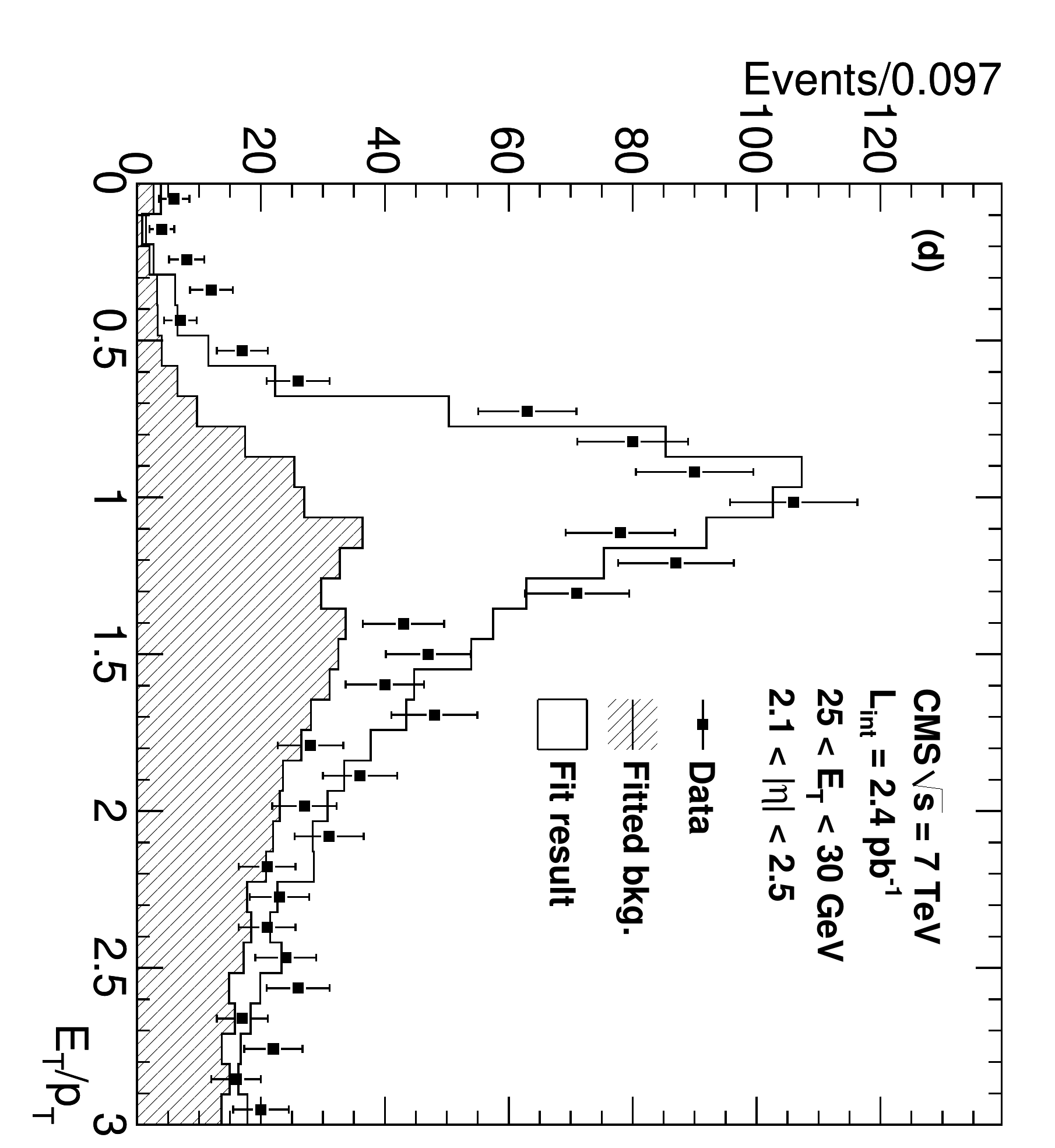}
\caption{Measured $\etgamma/\ptgamma$ distributions for converted photon
candidates with $\etgamma$ = 25--30\GeV, in the four $\etagamma$ regions considered.
The binned extended maximum likelihood fit result (open histogram) is overlaid
in each plot.
The fitted component from background is shown by hatched histograms.}
\label{fig:fit_bin1}
\end{figure*}

\begin{figure*}[htbH]
\centering
\includegraphics[width=\figwid]{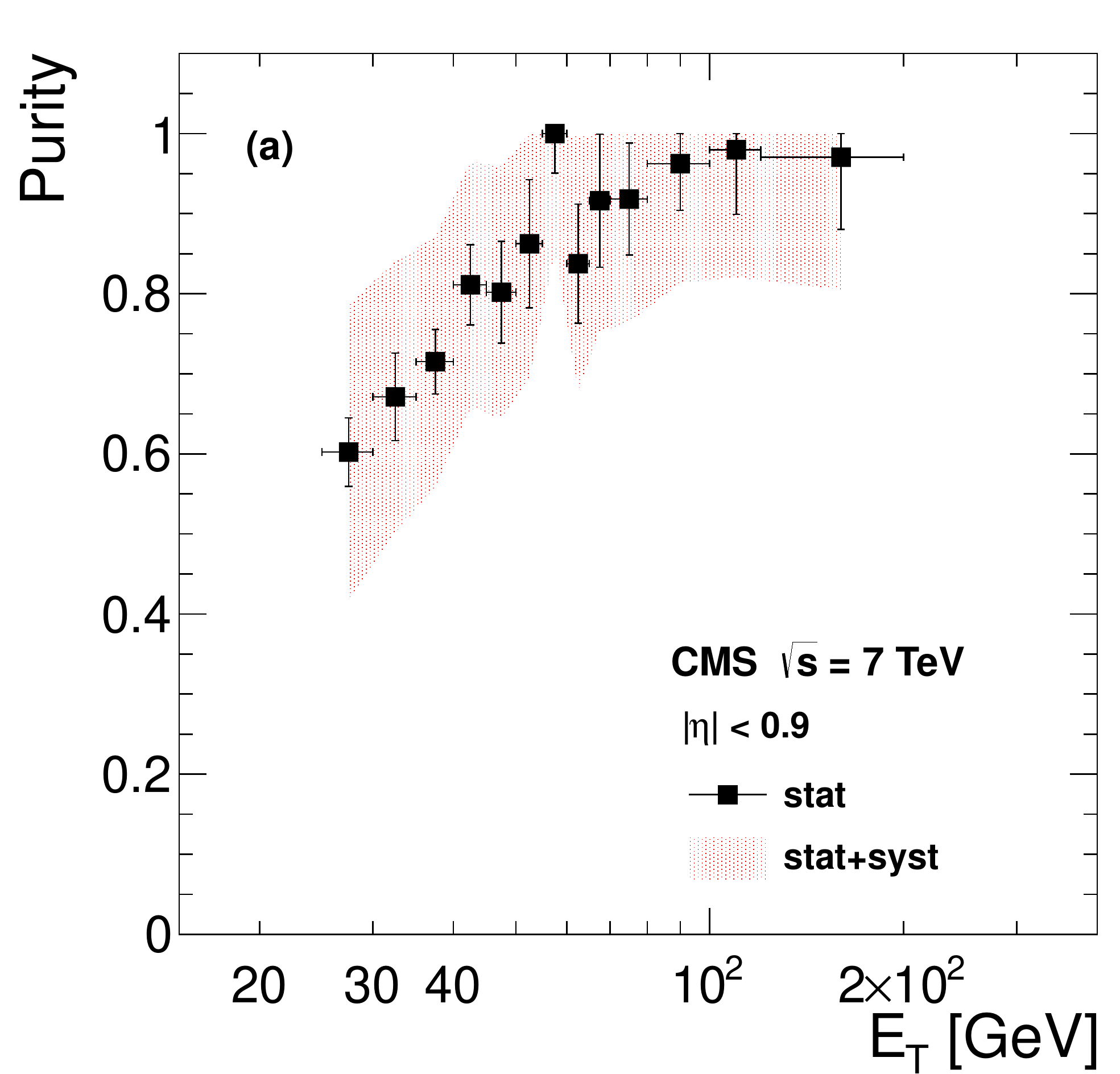}
\includegraphics[width=\figwid]{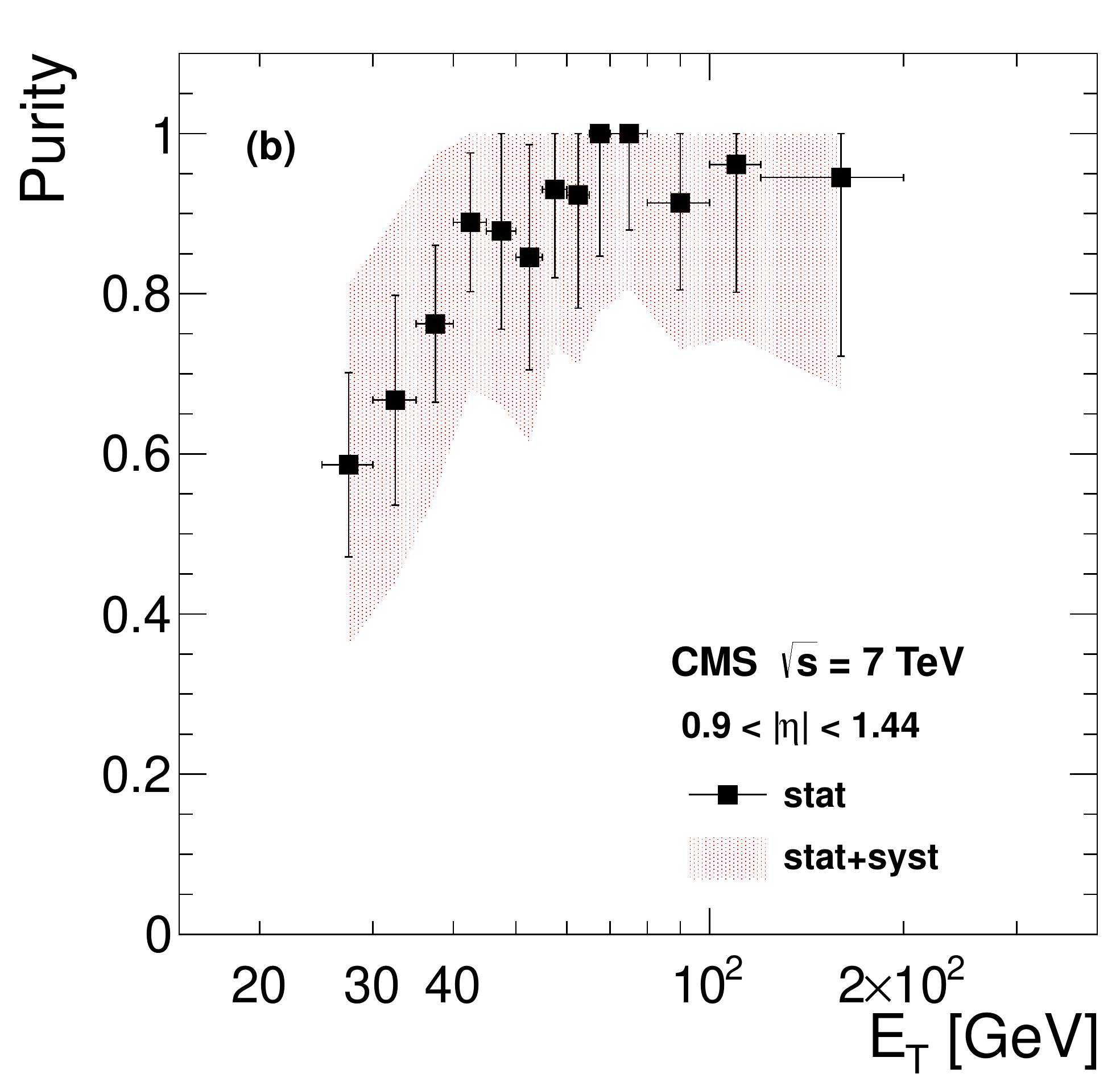}\\
\includegraphics[width=\figwid]{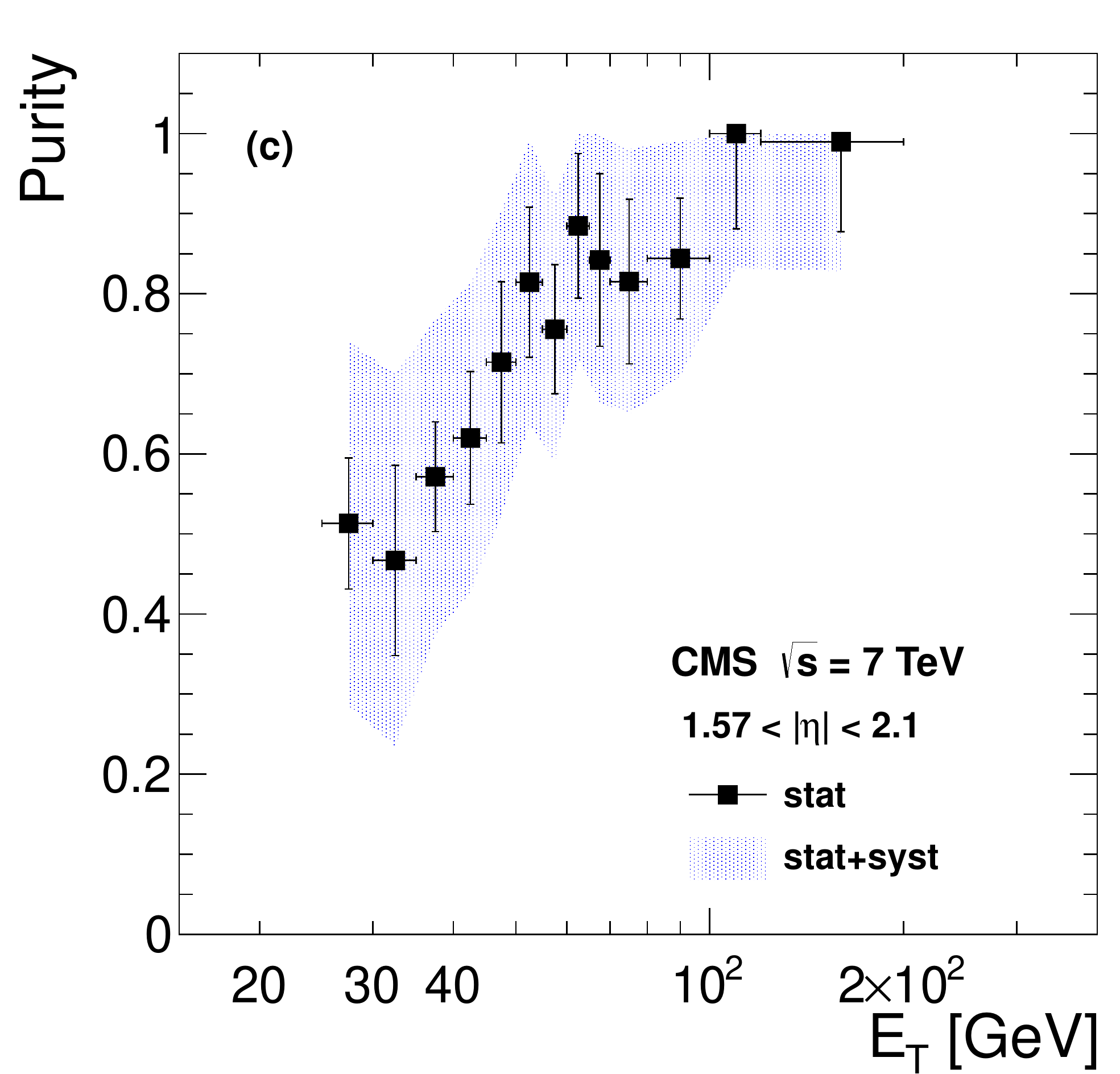}
\includegraphics[width=\figwid]{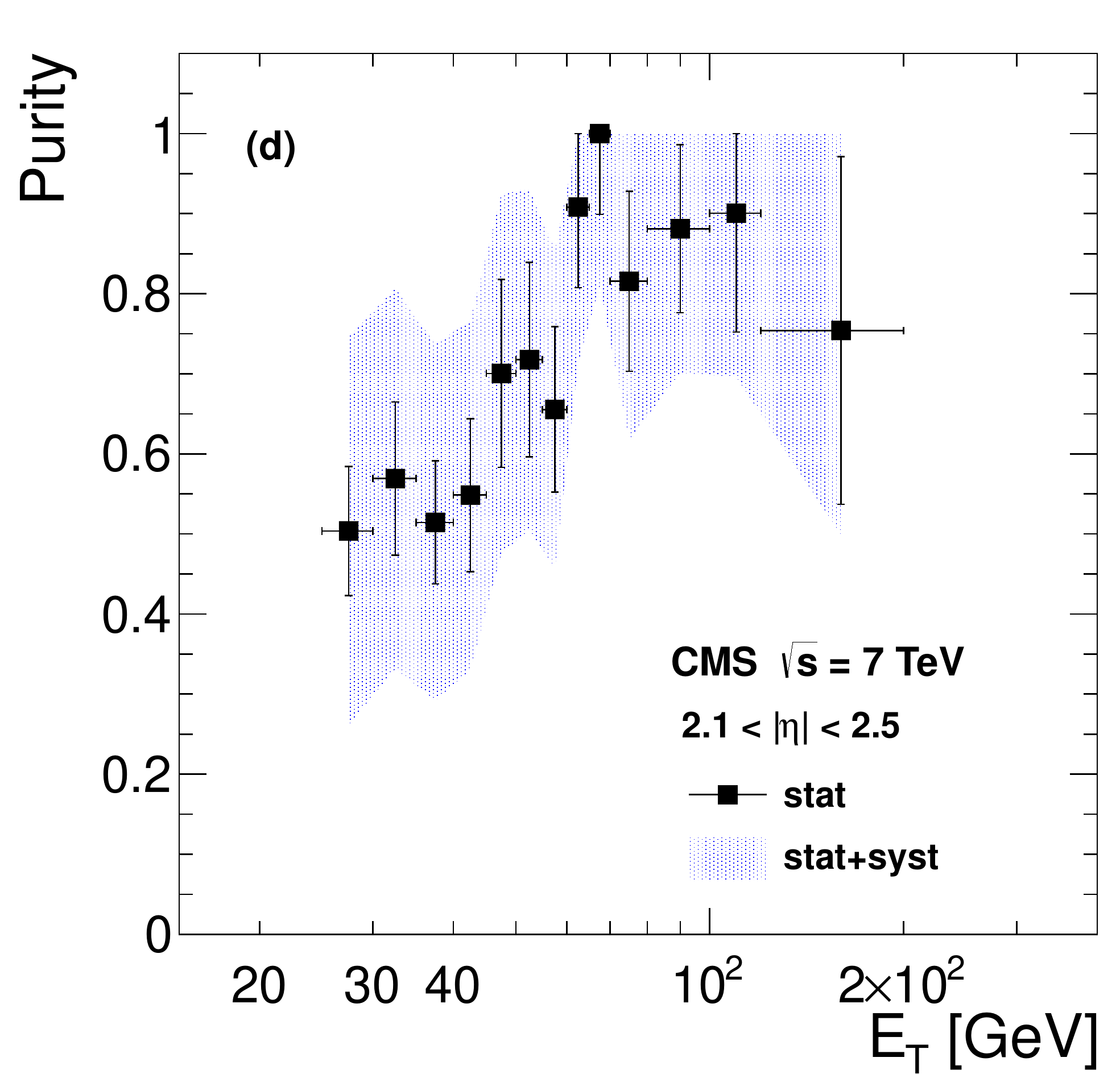}
\caption{Measured signal purity for $0<\etgamma/\ptgamma < 3$
with the photon conversion method in the four $\etagamma$ regions.
 The vertical error bars show the statistical uncertainties, while the
 shaded areas show the statistical and systematic
 uncertainties added in quadrature.
 Estimation of the systematic uncertainties is discussed in Section~\ref{sec:sys}.
 }\label{fig:convPurity}
\end{figure*}

An example of the fit to data is shown in Fig.~\ref{fig:fit_bin1} and
the photon yield $N^{\cPgg}$ for each \etgamma\ and \etagamma\ interval
is listed in Table~\ref{tab:yield}. Due to the lack of converted photon
candidates at large \etgamma, photon conversion results are measured for
$\etgamma < 200\GeV$. Figure~\ref{fig:convPurity} shows the measured signal
purity, defined as the estimated fraction of true prompt photons over all
reconstructed photons that satisfy the selection criteria. The signal purity
clearly increases with photon transverse energy, as expected from simulation
studies.

\begin{table*}[htbp]
\caption{Measured signal yield $N^{\cPgg}$ from the photon conversion method.
 The uncertainty on the yield is the statistical
uncertainty from the extended maximum likelihood fit.
\label{tab:yield}}
\begin{center}
\begin{tabular}{l c c c c} \hline \hline
 $\etgamma$ (\!\GeV)  & $|\etagamma|<0.9$ & $0.9<|\etagamma|<1.44$ & $1.57<|\etagamma|<2.1$ & $2.1<|\etagamma|<2.5$ \\ \hline
25--30 & 1254 $\pm$ 44 & 275 $\pm$ 26 & 661 $\pm$ 43 & 577 $\pm$ 40 \\
30--35 & 648 $\pm$ 31 & 157 $\pm$ 26 & 280 $\pm$ 31 & 298 $\pm$ 29 \\
35--40 & 1126 $\pm$ 40 & 262 $\pm$ 40 & 618 $\pm$ 39 & 446 $\pm$ 36 \\
40--45 & 711 $\pm$ 49 & 197 $\pm$ 21 & 362 $\pm$ 31 & 268 $\pm$ 26 \\
45--50 & 436 $\pm$ 35 & 115 $\pm$ 13 & 235 $\pm$ 40 & 170 $\pm$ 35 \\
50--55 & 262 $\pm$ 27 & 75 $\pm$ 10 & 183 $\pm$ 26 & 114 $\pm$ 18 \\
55--60   &  444 $\pm$ 27 & 101 $\pm$ 8 &  241 $\pm$ 31 &  142 $\pm$ 17 \\
60--65   &  255 $\pm$ 22 &  56 $\pm$ 5 &  159 $\pm$ 15 &  119 $\pm$ 12 \\
65--70   &  181 $\pm$ 13 &  41 $\pm$ 6 &  104 $\pm$ 13 &   89 $\pm$  8 \\
70--80   &  254 $\pm$ 18 &  73 $\pm$ 9 &  130 $\pm$ 18 &   93 $\pm$ 14 \\
80--100  &  437 $\pm$ 19 &  98 $\pm$ 8 &  231 $\pm$ 26 &  122 $\pm$ 15 \\
100--120 &  177 $\pm$ 7  &  42 $\pm$ 3 &   61 $\pm$  8 &   41 $\pm$  5 \\
120--200 &  134 $\pm$ 6  &  22 $\pm$ 2 &   65 $\pm$  3 &   22 $\pm$  5 \\
\hline \hline
\end{tabular}
\end{center}
\end{table*}

\subsection{Isolation method \label{sec:combIso}}

Photons are required to satisfy the signal-region preselection
criteria listed in Table~\ref{tab:preCuts}.
Background from electrons is suppressed by requiring the absence of
a short track segment, built from either two or three hits in the silicon
pixel detector, consistent with an electron track matching the observed
location and energy of the photon candidate (pixel veto requirement).
The signal and background component distributions are parametrised with
analytic functions. For the signal component, a convolution of
exponential and Gaussian functions is used:
\begin{equation}
\label{eq:sigPDF}
S(x)=\exp(ax)\otimes {\rm Gauss}(\mu, \sigma, x),
\end{equation}
while for the background the threshold function is used:
\begin{equation}
\label{eq:bkgPDF}
B(x)=(1-p_1 (x-p_0))^{p_2} \times (1-e^{p_3 (x-p_0)}).
\end{equation}
Here, $x$ is the $\Iso$ variable.
Other parametrisations give a much larger $\chi^2$ and fail to describe the
observed $\Iso$ distributions.
Using these parametrisations, an unbinned extended
maximum likelihood fit to the measured \Iso\ distribution is performed
for the region $-1<x<20\GeV$, with the likelihood defined as
\[
\label{eq:ExtML}
-\ln{L}=N_s+N_b - \sum^N_{i=1} \ln(N_s\mathcal{P}_s^i + N_b\mathcal{P}_b^i),
\]
where $N_s$ and $N_b$ are the expected signal and background yields, $N$ is
the number of observed photon candidates, and
$\mathcal{P}_s^i$ and $\mathcal{P}_b^i$ are the signal and background
probability density functions evaluated with the $\Iso$ of photon
candidate $i$. The signal and background probability density functions
are obtained by normalising the integrals of $S(x)$ and $B(x)$ to unity
in the fit range, respectively:
\[
\mathcal{P}_s^i = \frac{1}{\int_{-1}^{20} S(x)\mathrm{d}x} S(x^i),
\]
and
\[
\mathcal{P}_b^i = \frac{1}{\int_{-1}^{20} B(x)\mathrm{d}x} B(x^i),
\]
where $x^i$ is the measured $\Iso$ of photon candidate $i$.

While fitting the observed $\Iso$ distributions in data, the values of the
shape parameters in $\mathcal{P}_s$ and $\mathcal{P}_b$ are not fixed.
The two signal shape parameters $\mu$ and $\sigma$ in Eq.~(\ref{eq:sigPDF})
and two background shape parameters $p_1$ and $p_2$ in Eq.~(\ref{eq:bkgPDF})
are determined from the fit to data directly, while
the exponential tail of the signal $a$, the background turn-on power
$p_3$, and the background starting point $p_0$ are constrained in the fit.

The constrained values of parameter $a$ and parameters $p_0$ and $p_3$ are
obtained first by fitting the simulated signal events with the
parametrisation in Eq.~(\ref{eq:sigPDF}) and simulated background events with
the parametrisation in Eq.~(\ref{eq:bkgPDF}), respectively. Then, the
constrained values are further corrected with data-to-simulation scaling
factors. A difference between data and simulation is observed due to the
imperfect modelling of detector noise, the underlying event, pile-up, and
the hadronization process.

To derive the scaling factor for the parameter $a$, low-bremsstrahlung
electrons from $\Zz\to \EE$ decays are selected as described in
Ref.~\cite{ref:EGM-10-004}. The amount of bremsstrahlung is obtained from
the relative difference between the momentum measured at the last point
($p_\mathrm{out}$) on the electron track and the momentum measured at the
origin ($p_\mathrm{in}$). Here, ``low bremsstrahlung'' means that the ratio
$\left(p_{\mathrm{in}}-p_{\mathrm{out}}\right)/p_{\mathrm{in}}$ is less than
0.15. A fit to the electron $\Iso$ distribution is performed using the
parametrisation in Eq.~(\ref{eq:sigPDF}); the ratio of the value of $a$
obtained from electron data to that from electron simulation is taken as the
scaling factor for the parameter $a$ of the photon signal shape.

To derive the scaling factors for the parameters $p_0$ and $p_3$, a
background-enriched sample is selected with the sideband-region selection
criteria listed in Table~\ref{tab:preCuts}; the contamination of signal
in this sideband region is negligible. In this sideband region, photon
candidates satisfy the loose but fail the tight criterion on
$\sigma_{\eta\eta}$; the other requirements are the same as those for
the signal region.
Then, fits to the sideband-region $\Iso$ distributions in the data and
in the simulation, using the parametrisation in Eq.~(\ref{eq:bkgPDF}), are
performed to obtain the scaling factor.

\begin{figure*}[hbtp]
\centering
\includegraphics[width=\figwid]{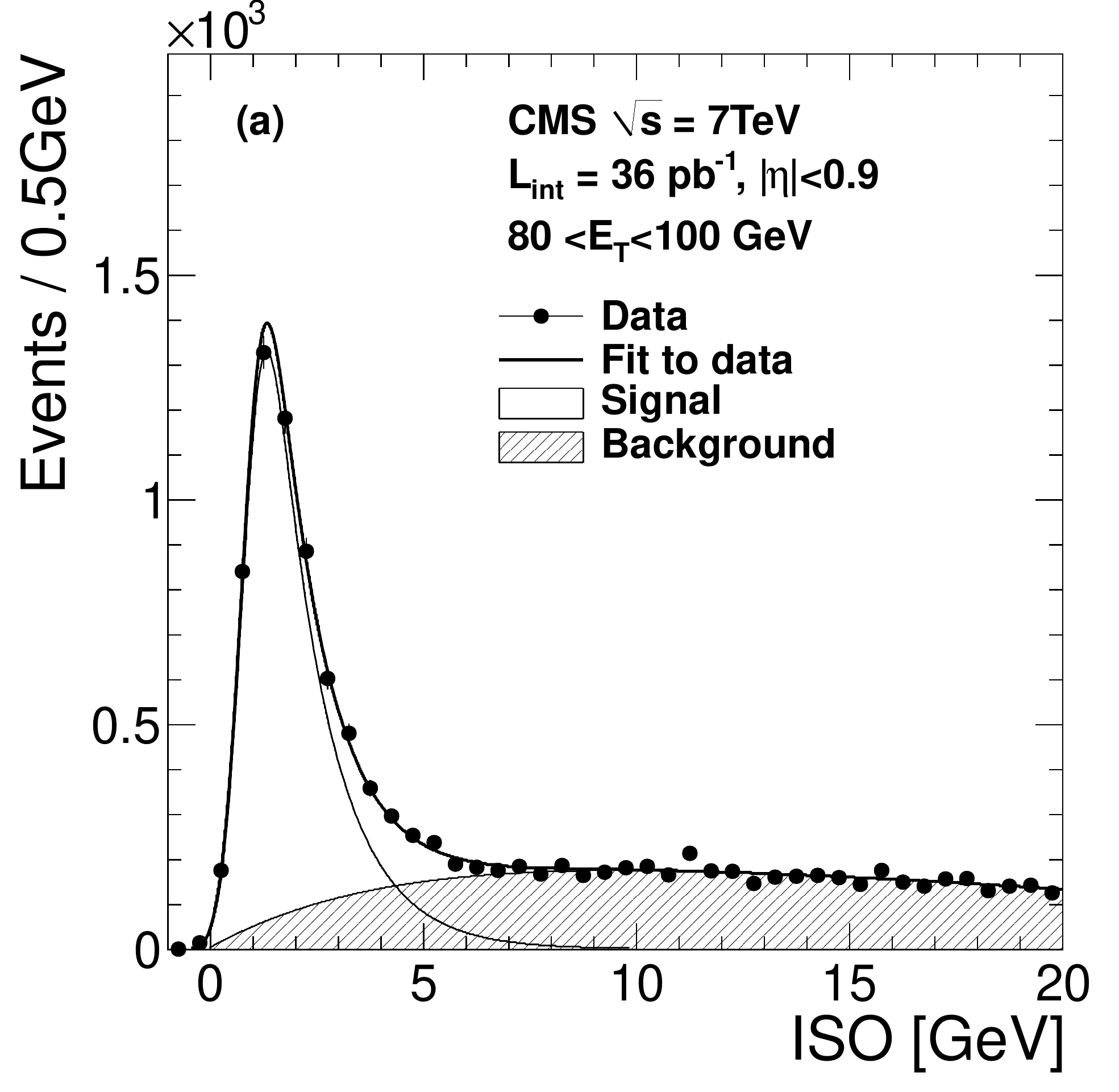}
\includegraphics[width=\figwid]{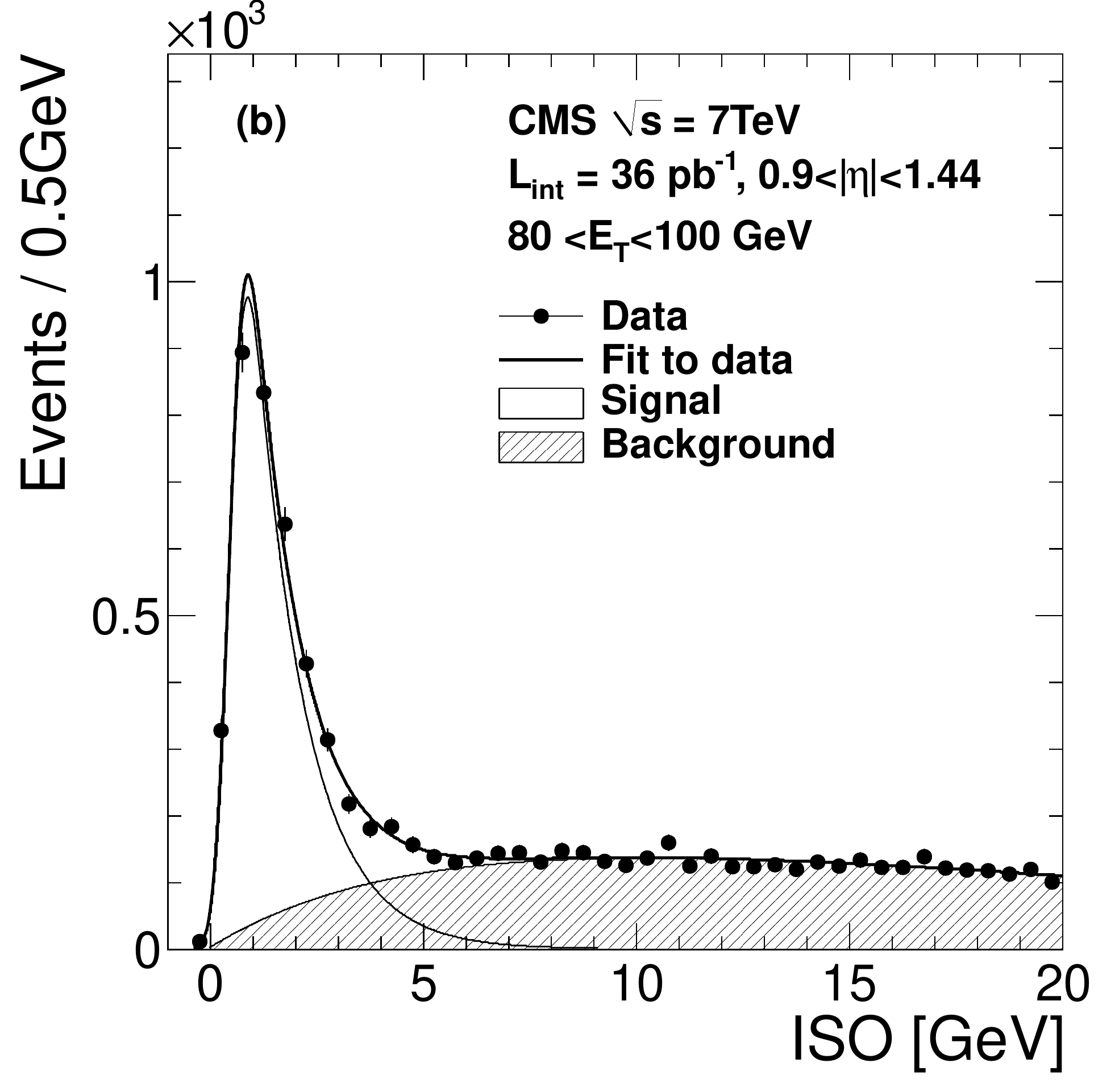}\\
\includegraphics[width=\figwid]{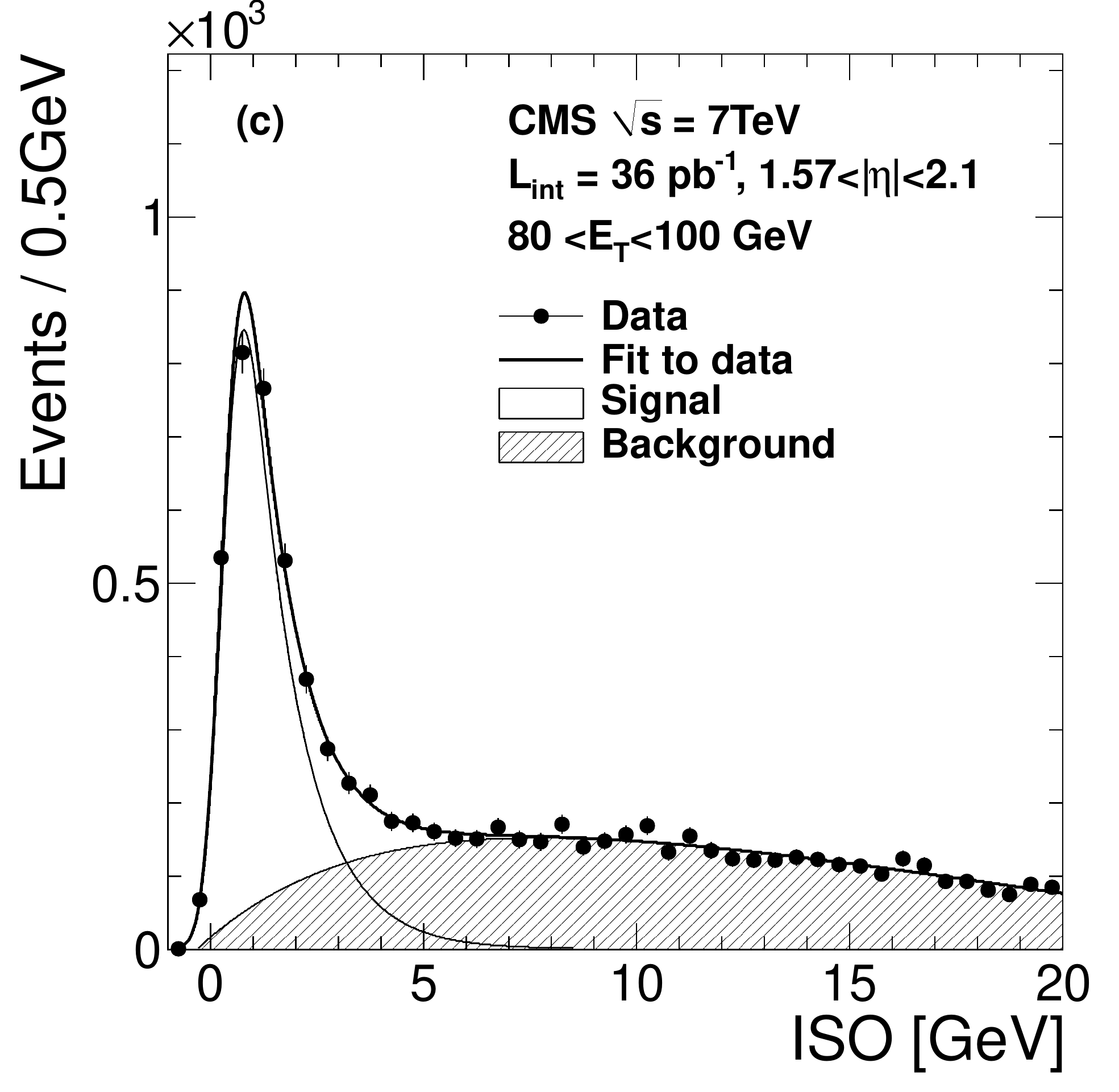}
\includegraphics[width=\figwid]{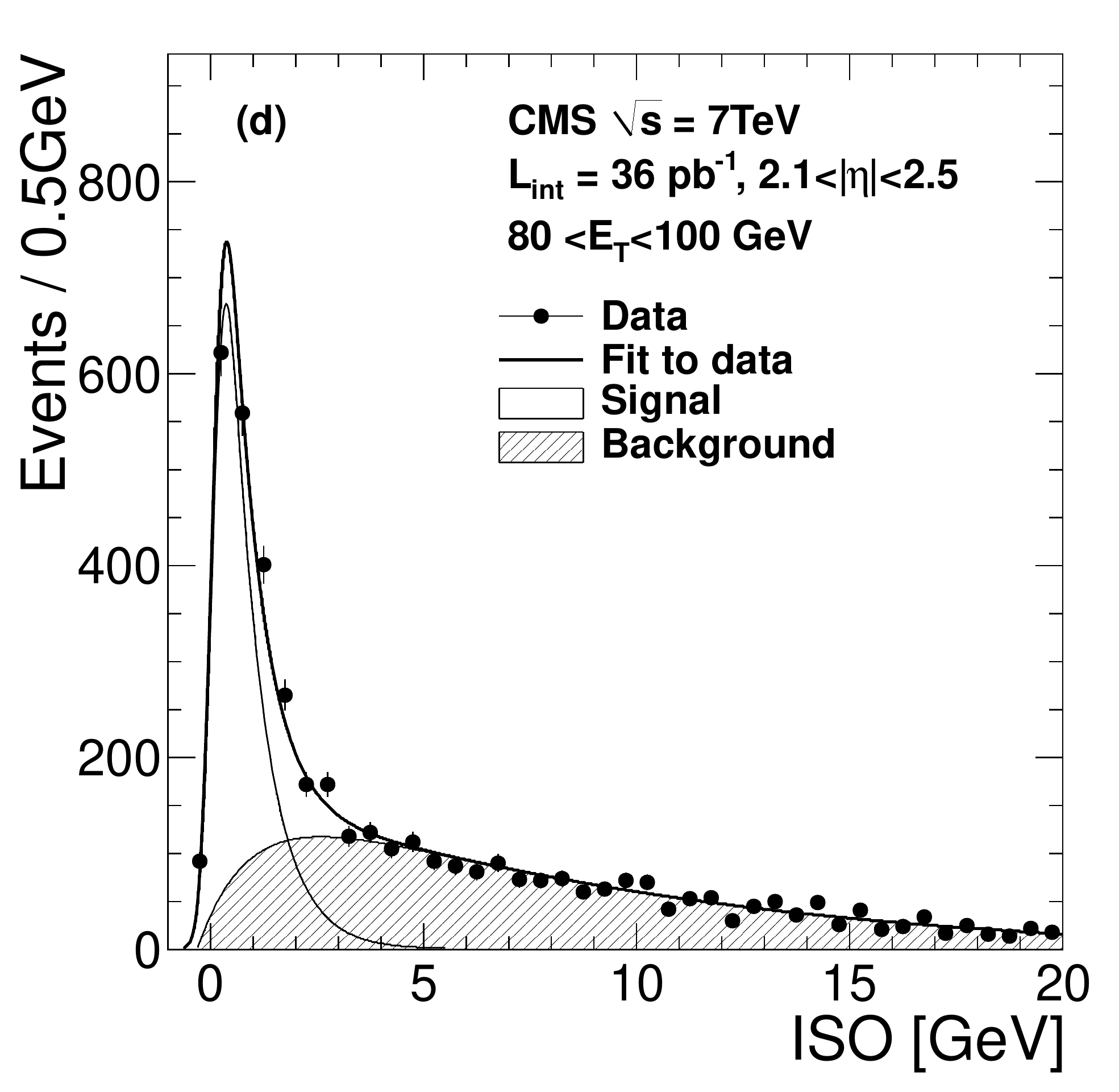}
\caption{Measured $\Iso$ distributions for candidates with
$\etgamma$ = 80--100\GeV. The unbinned maximum likelihood fit result
(solid line) is overlaid in each plot. The fitted signal and background
components are also shown. Imperfections of the fitting model are included as
part of the systematic uncertainties.}
\label{fig:iso_fit}
\end{figure*}

\begin{figure*}[htbpH]
\centering
\includegraphics[width=\figwid]{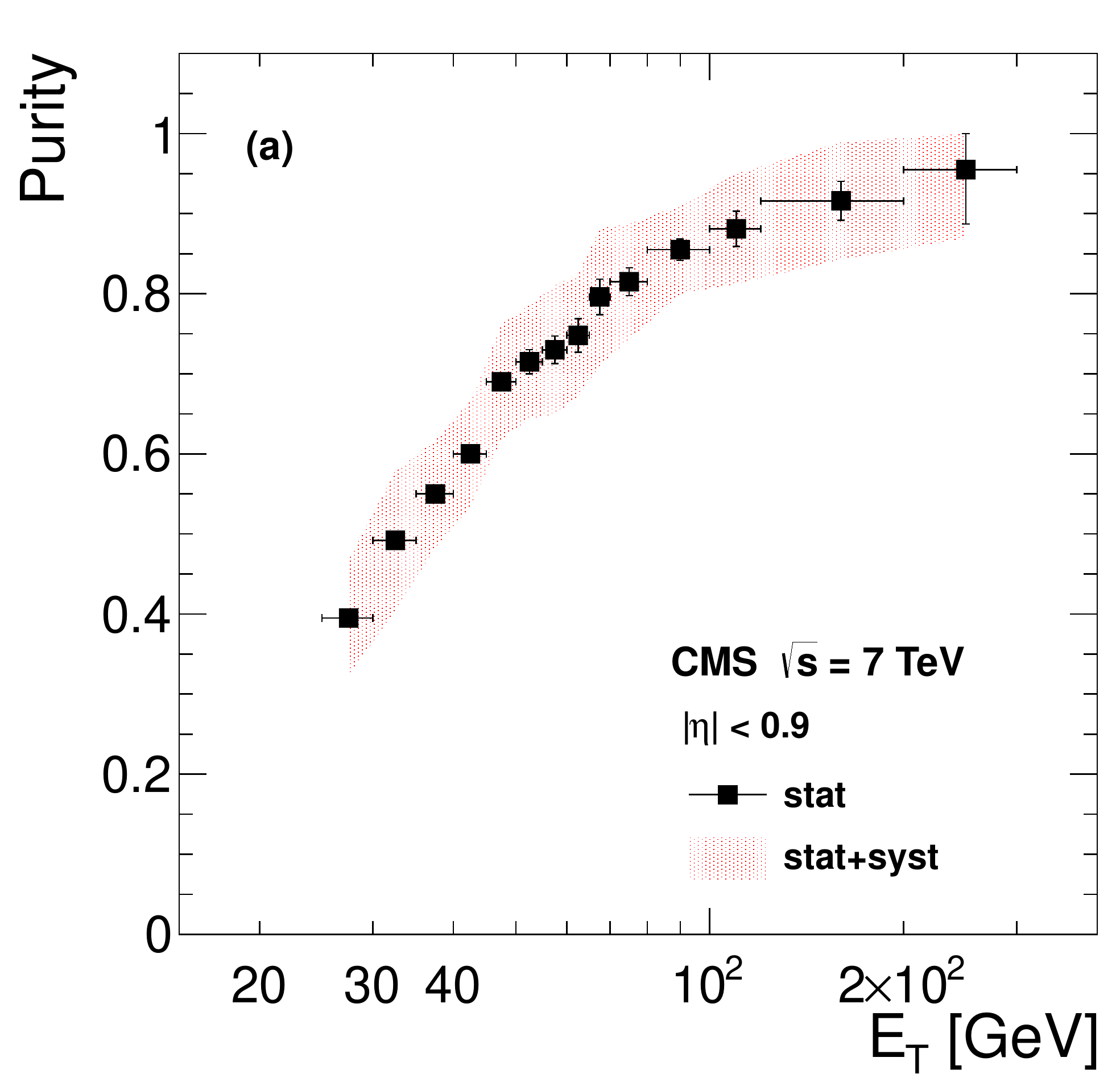}
\includegraphics[width=\figwid]{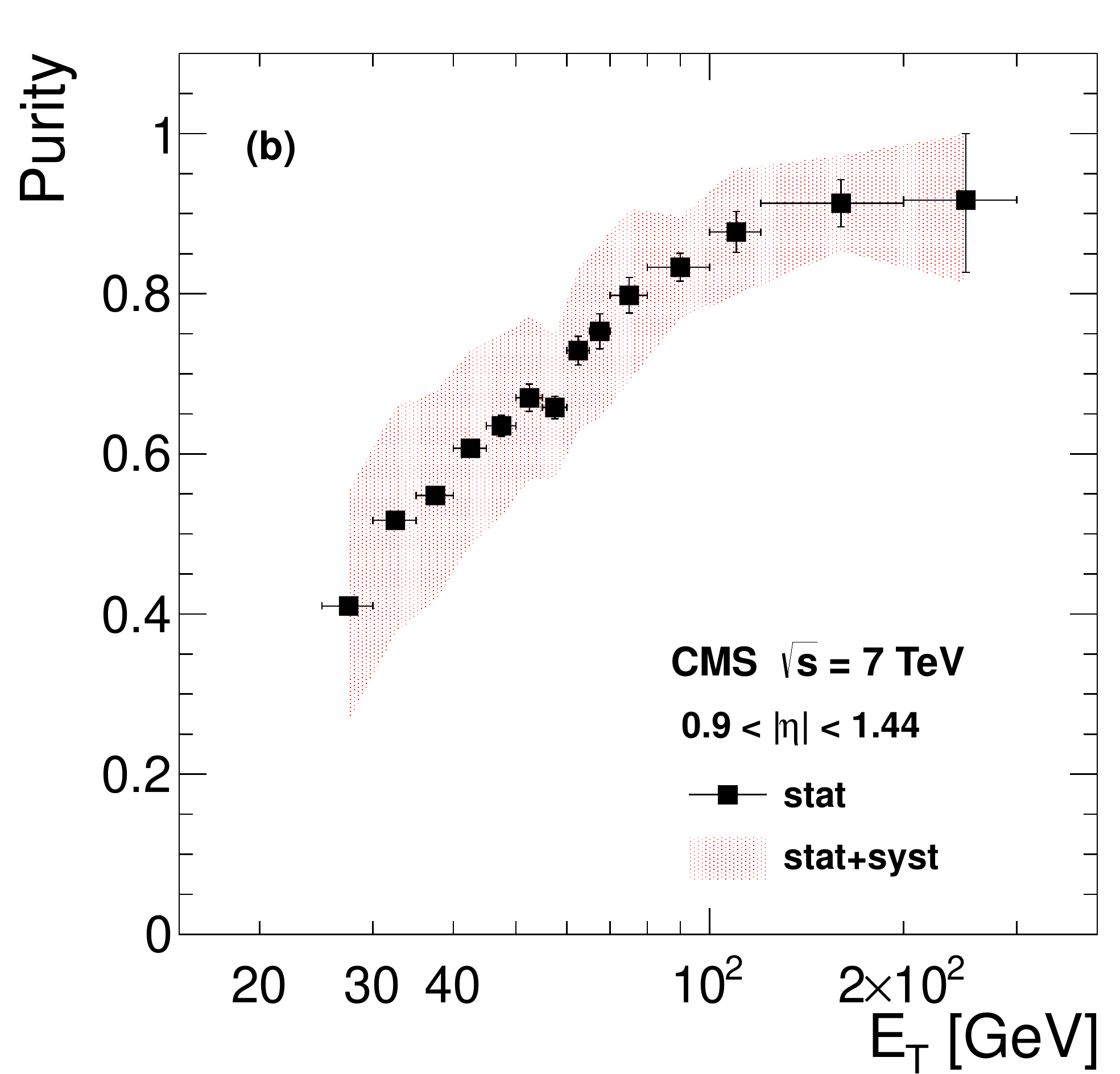}\\
\includegraphics[width=\figwid]{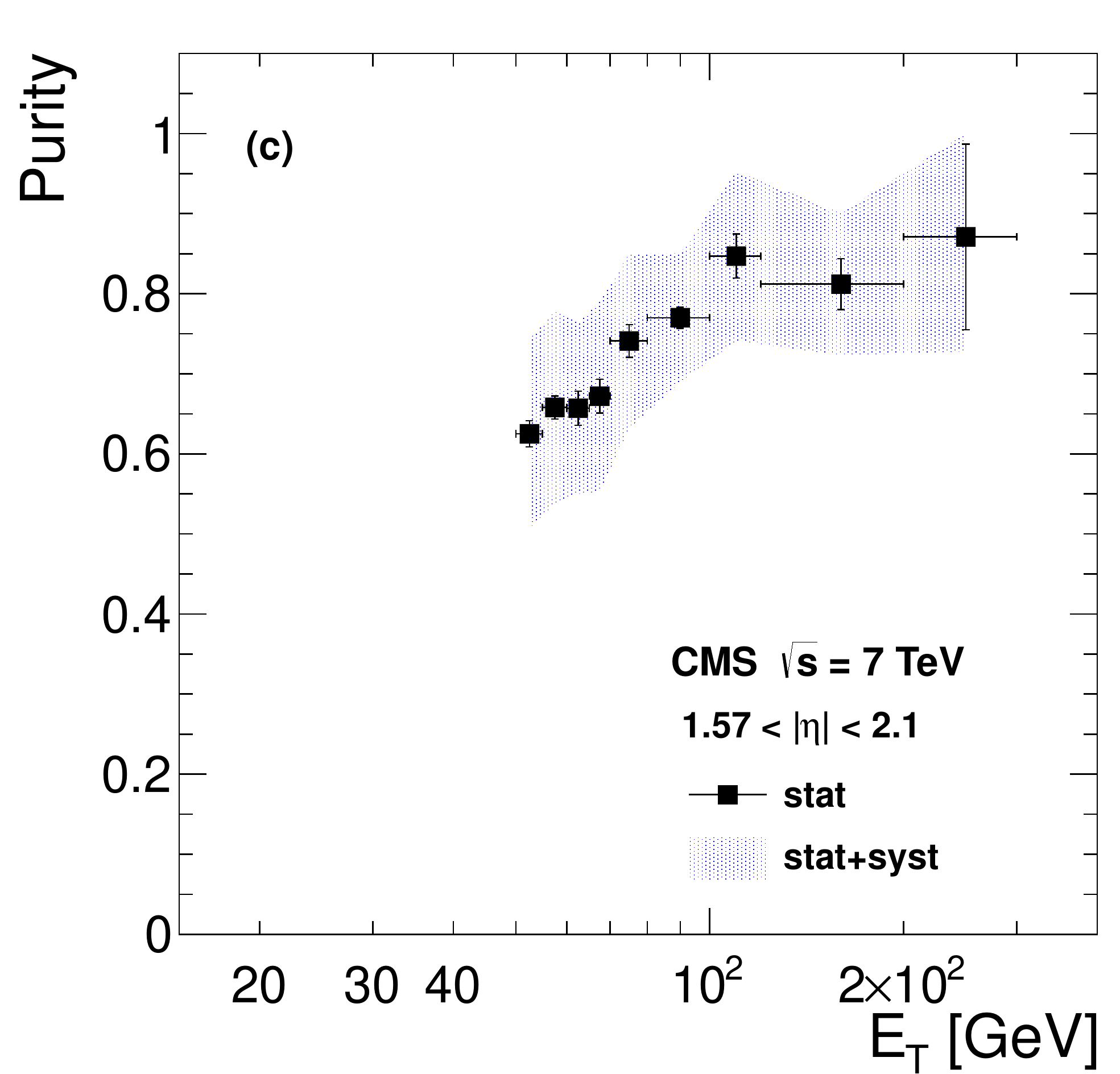}
\includegraphics[width=\figwid]{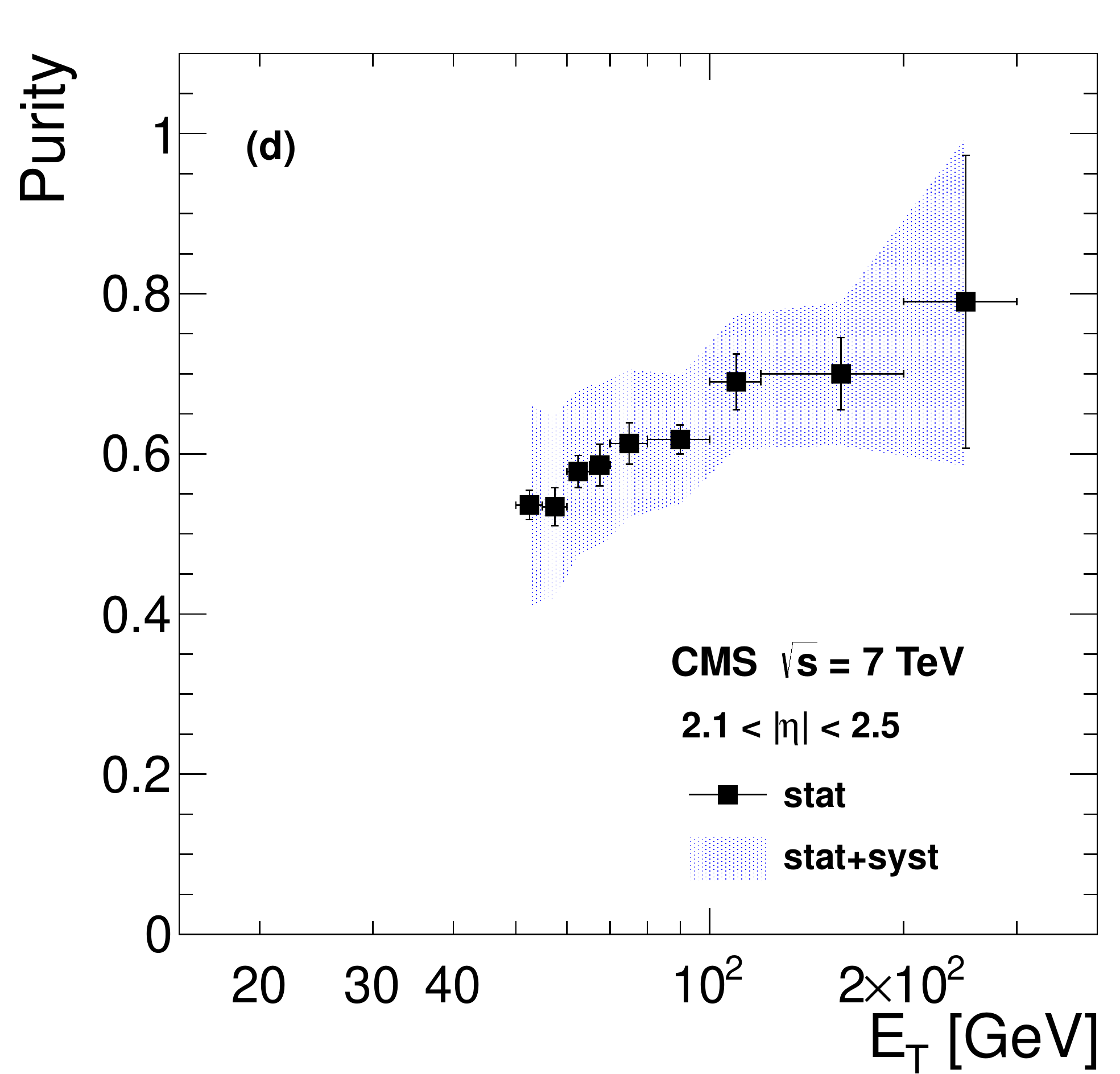}
\caption{Measured signal purity for $\Iso<5\GeV$ with the isolation method in the
 four $\etagamma$ regions. The purity for the bin $\etgamma$ = 300--400\GeV is
 not shown. The vertical error bars show the statistical uncertainties, while
 the shaded areas show the statistical and systematic uncertainties added in
 quadrature.
 Estimation of the systematic uncertainties is discussed in Section~\ref{sec:sys}.
 }
\label{fig:isoPurity}
\end{figure*}

Figure~\ref{fig:iso_fit} shows the result of the fit for
photons with $\etgamma$ = 80--100\GeV in the four $\etagamma$ intervals.
The value of $\Iso$ may be negative given that an average value is used
to subtract the contribution of detector noise in the computation of
$\ecaliso$ and $\hcaliso$ variables.
Table~\ref{tab:isoyield} lists the signal yields $N^{\cPgg}$ for each \et
and $\eta$ bin. The results for $\etgamma < 50\GeV$ in the endcaps are not
used in the measurement due to the large systematic uncertainties in the
modelling of the background shape. In order to minimise dependence on the
model of isolation, the signal yields are quoted for $\Iso < 5\GeV$.
Because the signal and background-enriched samples are small in the highest
$\etgamma$ bin (300--400\GeV), the
value of $N^{\cPgg}$ is obtained by counting the number of
observed photon candidates assuming 100\% purity, instead of performing
the fit. 
Such an assumption is justified by the fact that purity increases with
$\etgamma$, as shown in Fig.~\ref{fig:isoPurity}.

\begin{table*}[htbp]
\caption{Measured signal yield $N^{\cPgg}$ from the isolation method.
 The uncertainty on the yield is the statistical uncertainty from the extended
maximum likelihood fit.}
\begin{center}
  \begin{tabular}{ l c c c c} \hline \hline
   $\etgamma$ (\!\GeV) & $ |\etagamma|<0.9$ & $0.9<|\etagamma|<1.44$ & $1.57<|\etagamma|<2.1$ & $2.1<|\etagamma|<2.5$\\\hline
    25--30                  &  15951 $\pm$ 209     &  12088 $\pm$ 165   & -- & -- \\
    30--35                  &  8193  $\pm$ 151     &  5977  $\pm$ 101   & -- & -- \\
    35--40                  &  14813 $\pm$ 179     &  10384 $\pm$ 131   & -- & -- \\
    40--45                  &  8568  $\pm$ 121     &  5790  $\pm$ 94    & -- & -- \\
    45--50                  &  5548  $\pm$ 92      &  3425  $\pm$ 72    & -- & --\\
    50--55                  &  3400  $\pm$ 71      &  2138  $\pm$ 54    &  2154	$\pm$ 56 &  1298 $\pm$ 44  \\
    55--60                  &  4906  $\pm$ 115     &  3067  $\pm$ 67    &  3155	$\pm$ 69 &  1747 $\pm$ 77  \\
    60--65                  &  3280  $\pm$ 92      &  2143  $\pm$ 52    &  2015	$\pm$ 66 &  1209 $\pm$ 42  \\
    65--70                  &  2397  $\pm$ 67      &  1521  $\pm$ 44    &  1378	$\pm$ 44 &  822  $\pm$ 36  \\
    70--80                  &  3013  $\pm$ 64      &  1928  $\pm$ 54    &  1812	$\pm$ 50 &  1042 $\pm$ 44  \\
    80--100                 &  5487  $\pm$ 85      &  3489  $\pm$ 73    &  3193	$\pm$ 54 &  1679 $\pm$ 49  \\
    100--120                &  2128  $\pm$ 53      &  1397  $\pm$ 41    &  1210	$\pm$ 39 &  572  $\pm$ 29  \\
    120--200                &  1842  $\pm$ 49      &  1111  $\pm$ 36    &  887	$\pm$ 35 &  396  $\pm$ 25  \\
    200--300                &  217   $\pm$ 15      &  121   $\pm$ 12    &  81	$\pm$ 11 &  27   $\pm$ 6   \\
    300--400                &  44    $\pm$  7      &  26    $\pm$  5    &   8	$\pm$  3 &   1   $\pm$ 1   \\
  \hline\hline
  \end{tabular}
\label{tab:isoyield}
\end{center}
\end{table*}

\section{Efficiency Estimation \label{sec:eff}}
The selection efficiency can be factorised into several contributions, corresponding to
the different steps of the selection process, and can be expressed as
\[
  \epsilon = \epsilon_\mathrm{reco} \times
             \epsilon_\mathrm{id1}  \times
             \epsilon_\mathrm{trig} \times
             \epsilon_\mathrm{id2}.
\label{eq:eff}
\]

The reconstruction efficiency $\epsilon_\mathrm{reco}$ is defined as the ratio
of the number of true prompt photons that are reconstructed to the number of
true prompt photons that are generated with true $\etgamma$ and $\etagamma$
and have a generator-level isolation less than 5\GeV
(Section~\ref{sec:sample}). The value of $\epsilon_\mathrm{reco}$ is
99.8\% for all $\etgamma$ and $\etagamma$ bins and is determined from
simulated photon signal events.

The preselection efficiency $\epsilon_\mathrm{id1}$ is defined as the ratio
of the number of true prompt photons that are reconstructed and satisfy the
preselection requirements in Table~\ref{tab:preCuts} (with additional
requirement
of $\Iso < 5\GeV$ for the isolation method) to the number of true
prompt photons that are reconstructed. The value of $\epsilon_\mathrm{id1}$ is
determined from simulated photon signal events first and then multiplied by a
data-to-simulation scaling factor. To derive this data-to-simulation scaling
factor, a technique called ``tag-and-probe''~\cite{ref:EGM-07-001} that uses
electrons from $\Zz\to\EE$ decays is applied.
The simulation predicts a few percent difference in the efficiency
$\epsilon_\mathrm{id1}$
between photons and electrons; half of this difference is taken
as a systematic uncertainty. In addition, the scaling factor is measured
in various time periods that correspond to different average
numbers of pile-up events due to multiple $\Pp\Pp$ interactions in the same bunch
crossing. The envelope of the full variation, approximately 3--5\% depending on the selection criteria and photon $\etgamma$, is taken as the systematic
uncertainty. The derivation of the efficiency scaling factors is described in
more detail in Ref.~\cite{ref:EGM-10-006}.

The trigger efficiency $\epsilon_\mathrm{trig}$ is measured with the
tag-and-probe method directly from data and defined with respect to the number
of reconstructed electrons satisfying the preselection criteria.
The $\epsilon_\mathrm{trig}$ is measured to be $(99.8\pm0.1)\%$ for the
barrel and $(99.0\pm0.7)\%$ for the endcaps.

The symbol $\epsilon_\mathrm{id2}$ represents the efficiency of the
pixel veto requirement for the isolation method while for the
photon conversion method it represents the product of three terms: conversion
probability, track reconstruction efficiency, and identification efficiency.
While $\epsilon_\mathrm{id1}$ and
$\epsilon_\mathrm{trig}$ are calibrated with electrons using the
``tag-and-probe'' technique, $\epsilon_\mathrm{id2}$ must be measured using a
different method as described in Sections~\ref{sec:effconv} and
~\ref{sec:effiso}.

\subsection{Photon conversion method \label{sec:effconv}}
The tag-and-probe scaling factor on $\epsilon_\mathrm{id1}$ is on average
$0.963\pm0.050$ for the barrel and $0.990\pm0.053$ for the endcap.
The uncertainty on the scaling factor is dominated by the
uncertainty associated with the number of pile-up events, the background
estimate underneath the $\Zz$ mass peak, and the difference between photons
and electrons observed in the simulation.

For the photon conversion method, $\epsilon_\mathrm{id2}$ cannot be measured
from the $\Zz\to\EE$ events and is measured with a different sample.
First, a sample is selected in data by applying the $H/E$ and
\texttt{$\sigma_{\eta \eta}$} requirements listed in Table~\ref{tab:preCuts}.
The $\Iso$ distribution of these selected candidates is used to extract the
yield, $N_1$, of signal photons with $\Iso <5\GeV$, using the signal shape
from Eq.~(\ref{eq:sigPDF}) with the background shape obtained from
the simulation.
Second, a subsample of these candidates is selected, which, in addition to
passing the shower shape selection, also have reconstructed conversion tracks
meeting the conversion identification criteria as discussed in
Section~\ref{sec:conv}.
The signal extraction is performed again on the conversion subsample
to obtain an estimate of the number of signal photons, $N_2$, that converted
and passed the conversion identification selection.
The ratio between the extracted number of signal events before and after
applying the conversion selection, $N_2/N_1$, is used as an estimate of the
value of $\epsilon_\mathrm{id2}$.

In simulation, the $\epsilon_\mathrm{id2}$ depends only weakly on the
$\etgamma$ of the photon, but varies strongly with  $\etagamma$.  Due to the
relatively small number of events in the conversion subsample, an average value
of the $\epsilon_\mathrm{id2}$ for each $\etagamma$ bin is extracted and
then corrected for the $\etgamma$ dependence observed in the simulation.
As a cross-check, photon identification criteria and the $\Iso$ fit range
are varied. The measured $\epsilon_\mathrm{id2}$ is found to be
independent of the choice of these parameters.

Figure~\ref{fig:eff}~(a) shows the total efficiency $\epsilon$ for the
photon conversion method, after taking into account the scaling factors,
as a function of photon \etgamma\ in the four \etagamma\ regions.
The value of $\epsilon$ for the photon conversion method is lower than
that for the isolation method because of the probability
for a photon to convert before reaching the CMS electromagnetic calorimeter
and the relatively small conversion reconstruction efficiency.
The conversion probability, estimated at the generator level, is
between 20\% and 70\% in the region of $|\etagamma| < 1.44$ and
between 65\% and 70\% in the region of $1.57 < |\etagamma| < 2.5$.
The efficiency in the region $0.9<|\etagamma|<1.44$ is lower than
in the other regions because this region covers the area of
transition between the tracker barrel and endcap where the largest amount of
material, due to cables and services, is located. This region is especially
challenging for electron and conversion reconstruction.
Uncertainties on the $\epsilon_\mathrm{id1}$, $\epsilon_\mathrm{id2}$, and
trigger efficiency $\epsilon_\mathrm{trig}$ are included as sources of
systematic uncertainty on the final cross section measurement in
Section~\ref{sec:sys}.

\subsection{Isolation method \label{sec:effiso}}
The data-to-simulation scaling factor on $\epsilon_\mathrm{id1}$ measured with
the tag-and-probe method varies from
$0.971\pm0.073$ to $0.955\pm0.032$ for the barrel and
from $0.998\pm0.056$ to $0.990\pm0.056$ for the endcaps,
as \etgamma\ increases from 20\GeV to 45\GeV.

In addition, as mentioned in Section~\ref{sec:combIso}, a pixel veto
requirement is applied in the isolation method to suppress the contribution of
electron background.
The efficiency of the pixel veto requirement is estimated with
the photons from the final-state radiation of muons in $\Zz$ decays, \ie
$\Zz\to\MM\cPgg$ events.
The algorithm used in the pixel veto requirement may be affected by the
presence of nearby muon tracks leading to a false match to the photon.
To reduce this bias, events with photons that are close to the muons are
removed and the procedure is validated with simulation.
A data-to-simulation scaling factor is measured to be $0.996\pm0.013$ for the
barrel and $0.959\pm0.062$ for the endcaps.

\begin{figure}[hbtp]
\centering
\includegraphics[width=\figwid]{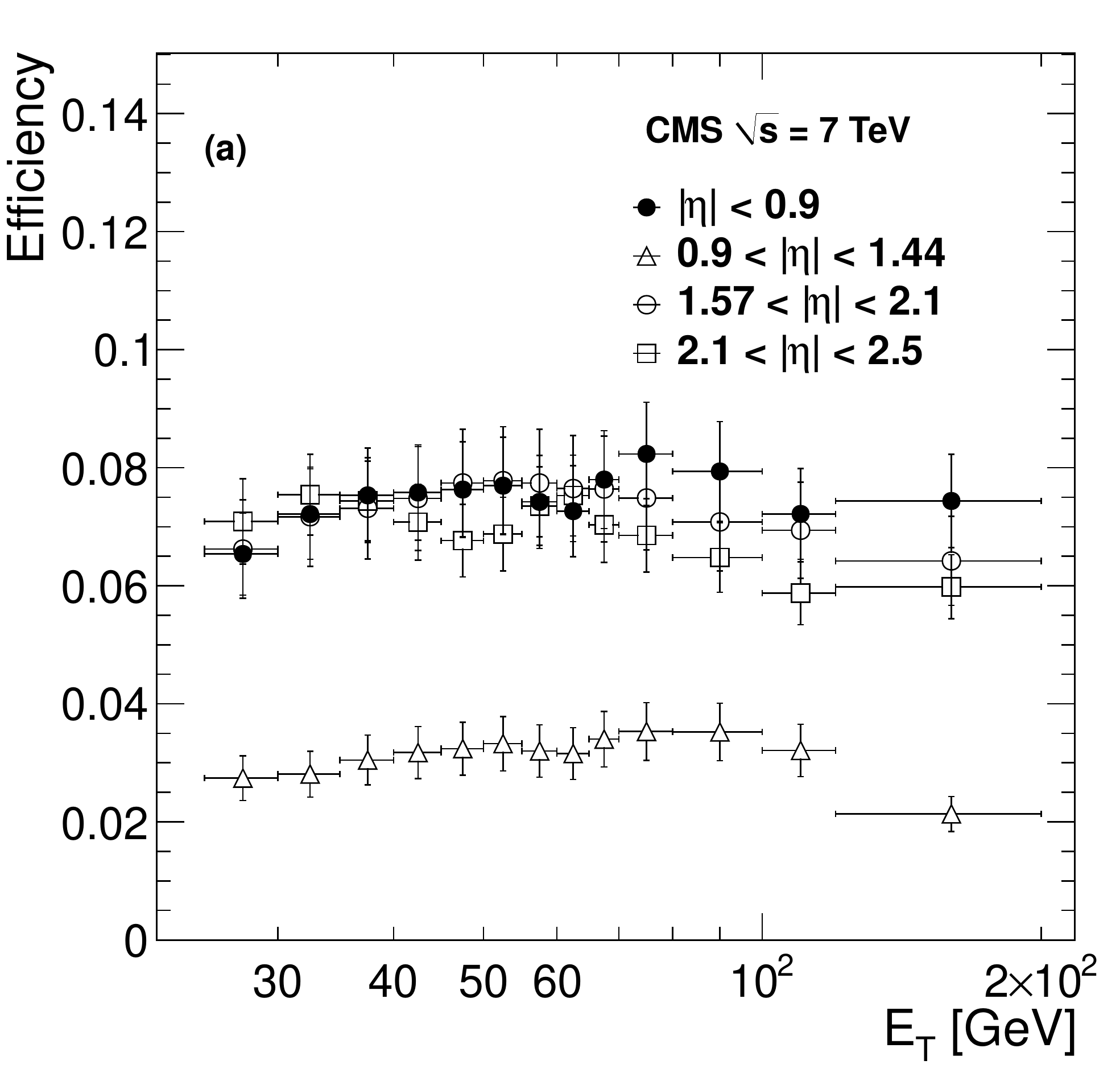}
\includegraphics[width=\figwid]{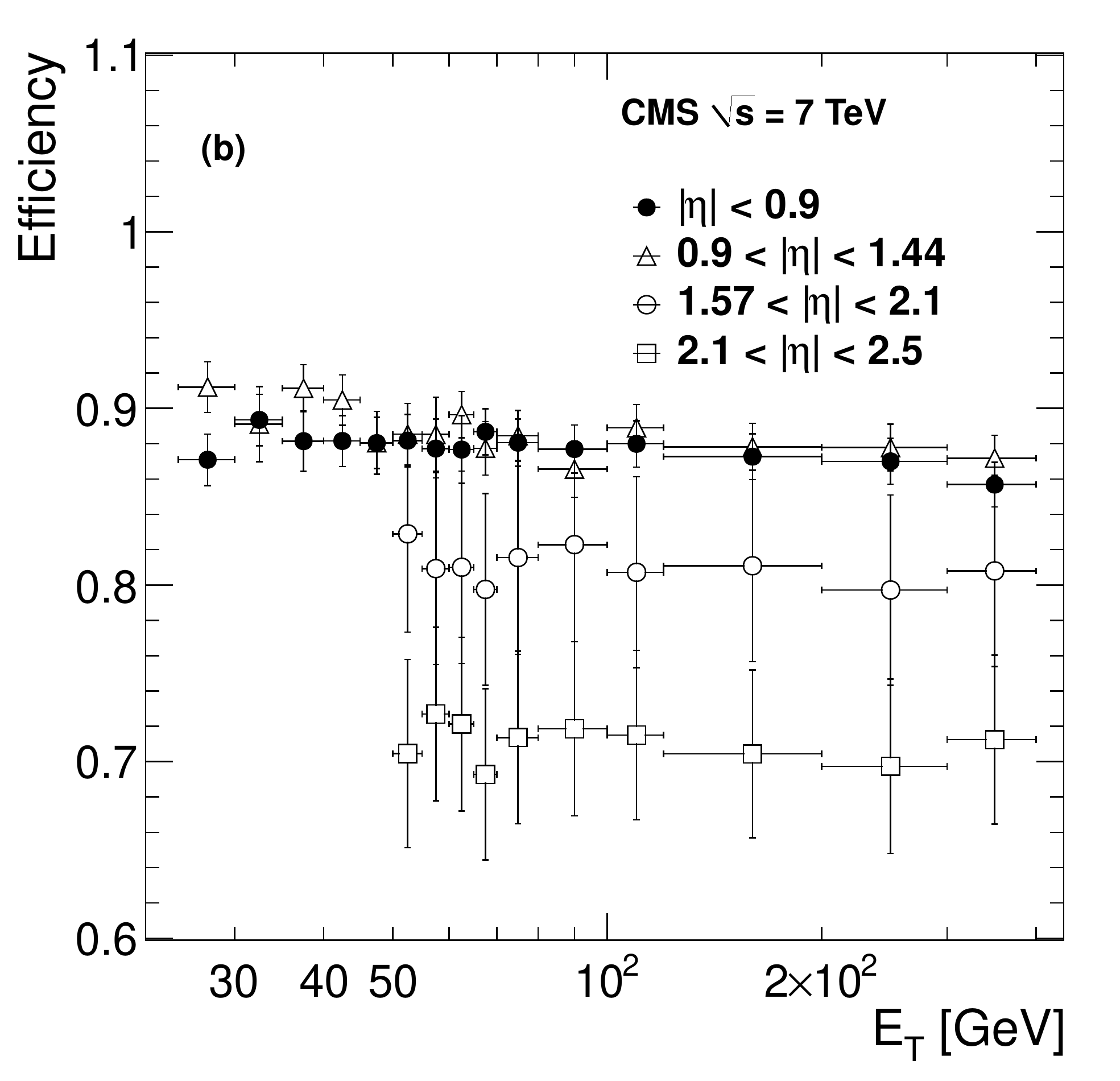}
\caption{Measured signal efficiency $\epsilon$ for the (a) photon conversion
and (b) isolation methods in the four \etagamma\ regions.
Data-to-simulation scaling factors have been applied. The error bars are dominated
by the systematic uncertainties and are 100\% correlated
between different \etgamma\ bins.}
\label{fig:eff}
\end{figure}

Figure~\ref{fig:eff}~(b) shows the total efficiency $\epsilon$ for the
isolation method, after taking into account the scaling factors, as a function
of photon \etgamma\ in the four \etagamma\ regions.
Uncertainties on the $\epsilon_\mathrm{id1}$, $\epsilon_\mathrm{id2}$,
and trigger efficiency $\epsilon_\mathrm{trig}$ are included as sources of
systematic uncertainty on the final cross section measurement in
Section~\ref{sec:sys}.

\section{Systematic Uncertainties \label{sec:sys}}

Table~\ref{tab:final_sys} summarises the systematic uncertainties on the
cross section in the four $\etagamma$ regions.
The major sources of systematic uncertainties include the uncertainties on
the shapes of the signal and background and photon identification efficiency.
Sections~\ref{sec:sysconv} and \ref{sec:sysiso}
describe the estimation of these dominant systematic uncertainties that
are specific to each method.

In addition to the $4\%$ overall uncertainty on the integrated luminosity,
uncertainties on the \ecal energy scale and trigger efficiency are 100\%
correlated between the photon conversion and isolation methods.
The uncertainty on the \ecal energy scale is estimated
from the \Zz mass peak positions to be 0.6\% for the barrel and 1.5\% for the
endcaps~\cite{ref:EGM-11-001}. The full analysis procedure is repeated by scaling
up and down the photon $\etgamma$ according to the uncertainty,
which results in an uncertainty of 4\% on the photon cross section. The 4\%
uncertainty is given by the statistical fluctuations in the yield rather than
the expected mean size of the effect.
The uncertainty on the trigger efficiency is limited by the available sample
of $\Zz\to\EE$ events.

For both methods, systematic uncertainties on the
signal and background shapes are obtained by pseudo-experiments.
The signal or background distribution is varied in the generated
pseudo-experiments according to the uncertainty on the shape
parameters. The result of each pseudo-experiment is then fitted
using the original fit model. The variation of the fitted yield is
assigned as the systematic uncertainty.
The uncertainties on the efficiency $\epsilon_\mathrm{id1}$ for both methods
and $\epsilon_\mathrm{id2}$ for the isolation method are dominated by the
limited number of $\Zz\to\EE$ and $\Zz\to\MM\cPgg$ events
available, the pile-up conditions, the background estimate, and the difference
between photons and electrons observed in simulation.

The uncertainties due to the bias introduced in the fitting procedure and the
amount of electron background result in a less than 6\% uncertainty on the
measured cross section.
The uncertainty due to imperfection of the fitting model is obtained from
pseudo-experiments by extracting the difference between observed
signal yields and the yields expected under the fitted model.
The electron background from $\Zz\to\EE$ decays is estimated
from the product of the integrated luminosity, the production cross section
measured in Ref.~\cite{ref:EWK-10-002}, and the efficiency from simulated
$\Zz$ events multiplied by a data-to-simulation efficiency scaling factor. The
contribution of electron background from $\PW\rightarrow \Pe\cPgn$ decays and
Drell--Yan processes is estimated following the same procedure. The total
contribution of electron background is less than 1\% and is taken as a
systematic uncertainty.

\begin{table*}[htbpH]
  \begin{center}
\caption{Systematic uncertainties expressed in percent for each source in the
four $\etagamma$ regions. The ranges, when quoted, indicate the variation over
photon $\etgamma$. The unfolding correction is discussed in
Section~\ref{sec:results}.        \label{tab:final_sys}}
\begin{tabular}{l| c c c c} \hline \hline
  Source & $|\etagamma|<0.9$ & $0.9<|\etagamma|<1.44$ & $1.57<|\etagamma|<2.1$ & $2.1<|\etagamma|<2.5$ \\ \hline
 & \multicolumn{4}{c}{Common} \\ \hline
  Luminosity         & \multicolumn{2}{c}{4.0} & \multicolumn{2}{c}{4.0} \\
  Energy scale       & \multicolumn{2}{c}{4.0} & \multicolumn{2}{c}{4.0} \\
  Trigger efficiency & \multicolumn{2}{c}{0.1} & \multicolumn{2}{c}{0.7} \\ \hline
 & \multicolumn{4}{c}{Photon conversion method} \\ \hline
  Isolation efficiency & 5.2 & 5.2 & 5.4 & 5.4 \\
  Conversion efficiency & 11   & 11   & 8.9 & 8.9 \\
  Fit bias & 0--4.1 & 0--6.1 & 0.1--4.2 & 0--5.3 \\
  Signal shape & 1 & 2.3 & 3 & 3.1 \\
  Background shape & 0.1--4.8 & 4.1--5.9 & 0.3--14   & 6.2--15   \\
  Electron background & 0.01--0.1 & 0.02--0.2 & 0.05--1.1 & 0.03--0.8 \\
  Unfolding correction & 4.0   & 4.0   & 4.0   & 4.0  \\ \cline{2-5}
  Total     &14--18   & 14--20 & 12--21  & 13--23 \\
  \hline
 & \multicolumn{4}{c}{Isolation method} \\ \hline
  Efficiency  & 3.6--7.6 & 3.6--7.6 & 8.6 & 8.6 \\
  Fit bias & 0.1--2.9 &0.1--2.8 &0.1--4.0 & 1.1--4.7 \\
  Signal/background shape &1.8--13   &1.6--32   & 4.9--16  & 7.0--21   \\
  $N^{\cPgg}$ for $\etgamma$ = 300--400\GeV & 4.5 & 8.3 & 10   & 20   \\
  Electron background & $<0.1$ & $<0.1$ & $<0.1$ & $<0.1$ \\
  Unfolding correction & 2.0   & 2.0    & 2.0 & 2.0 \\ \cline{2-5}
  Total     & 3.8--18   & 3.9--35   & 8.7--18   & 10--23   \\
  \hline\hline
  \end{tabular}
\end{center}
\end{table*}

\subsection{Photon conversion method \label{sec:sysconv}}
A significant source of systematic uncertainty affecting the photon conversion
method is the possible mismodelling of the signal and background probability
density functions ($\mathcal{P}_s$ and $\mathcal{P}_b$).
To establish the size of this uncertainty, both
$\mathcal{P}_s$ and $\mathcal{P}_b$ are checked against the data.
For the signal distribution, possible differences in the $\etgamma/\ptgamma$
distribution are investigated by varying the peak position and the width of
the distribution. For each variation, the change in signal yield is computed
along with the $\chi^2$ probability of the fit to data. The weighted variance
of these varied signal yields is computed using the $\chi^2$ probability for
each variation as the weight, and is used to set the systematic uncertainty on
the signal shape.
For the background, an alternate $\mathcal{P}_b$ is extracted from the
observed $\etgamma/\ptgamma$ distribution in the sideband (background-enriched) region
defined in Table~\ref{tab:preCuts}. The extraction of the signal yield
is repeated using this value of $\mathcal{P}_b$ determined from data, and the
size of the difference from the central value is taken as the systematic
uncertainty due to background shape. The observed and simulated
$\etgamma/\ptgamma$ distributions in the sideband region are in good
agreement and the shapes of $\etgamma/\ptgamma$ distributions are found to
be insensitive to the number of pile-up events in the data.

The systematic uncertainty associated with $\epsilon_\mathrm{id2}$ is
another significant source of uncertainty. The uncertainty on
$\epsilon_\mathrm{id2}$ due to the use of the isolation method to extract the
numbers of signal candidates $N_1$ and $N_2$ (Section~\ref{sec:effconv}) is
estimated in several ways, including a comparison of $\Iso$ background shapes
from data sideband regions and simulation and a comparison of
$\epsilon_\mathrm{id2}$ extracted with different photon candidate selections.
The statistical uncertainty on the number of candidates used in
Section~\ref{sec:effconv} to measure the $\epsilon_\mathrm{id2}$ from data is
also included in the uncertainty.
Finally, the $\epsilon_\mathrm{id2}$ is recomputed by splitting the data
sample into statistically independent halves. The final uncertainty on the
$\epsilon_\mathrm{id2}$ is chosen to cover the differences seen under each of
these variations.

\begin{figure*}[htbH]
\centering
\includegraphics[width=\figwid,angle=90]{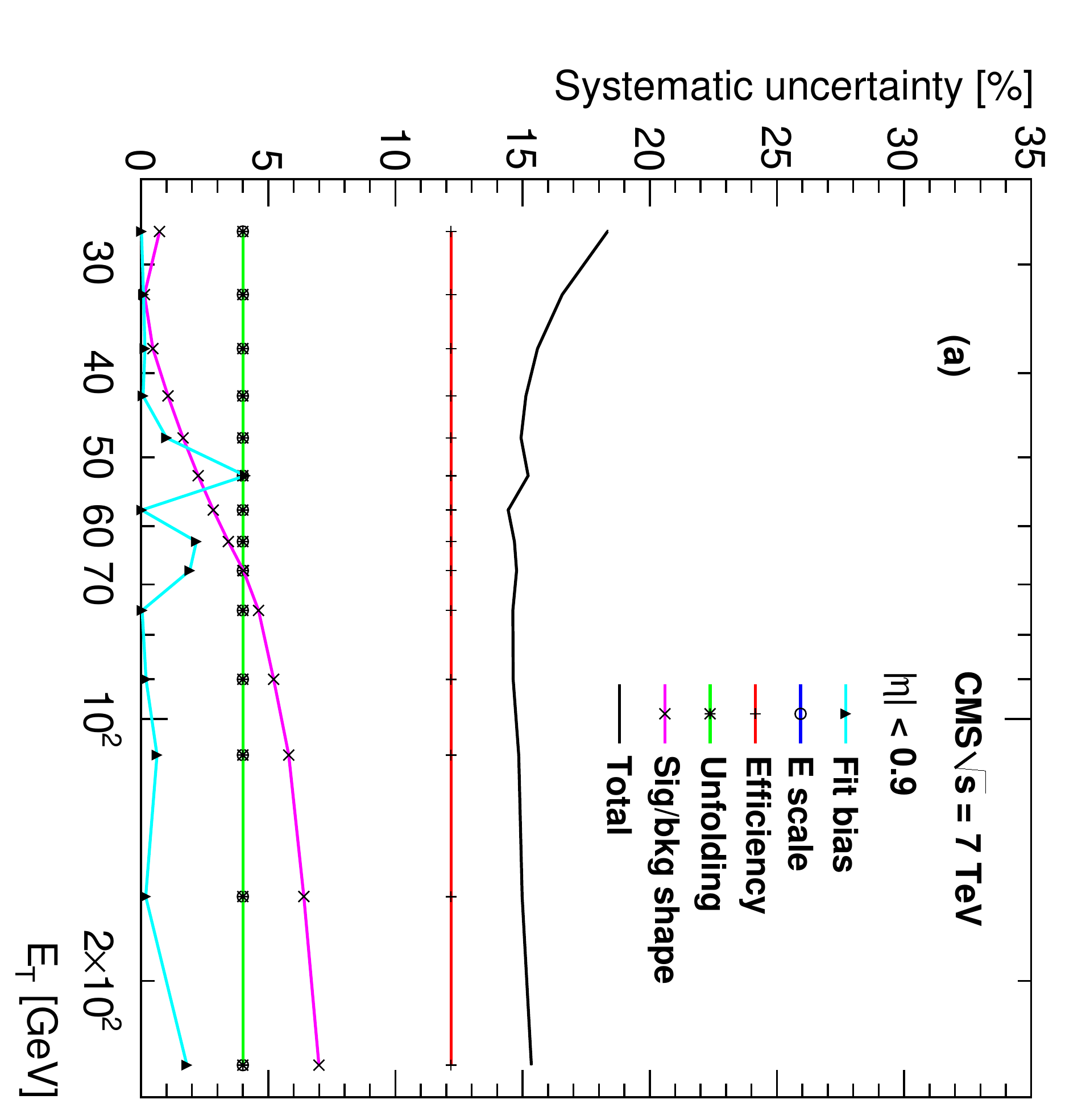}
\includegraphics[width=\figwid,angle=90]{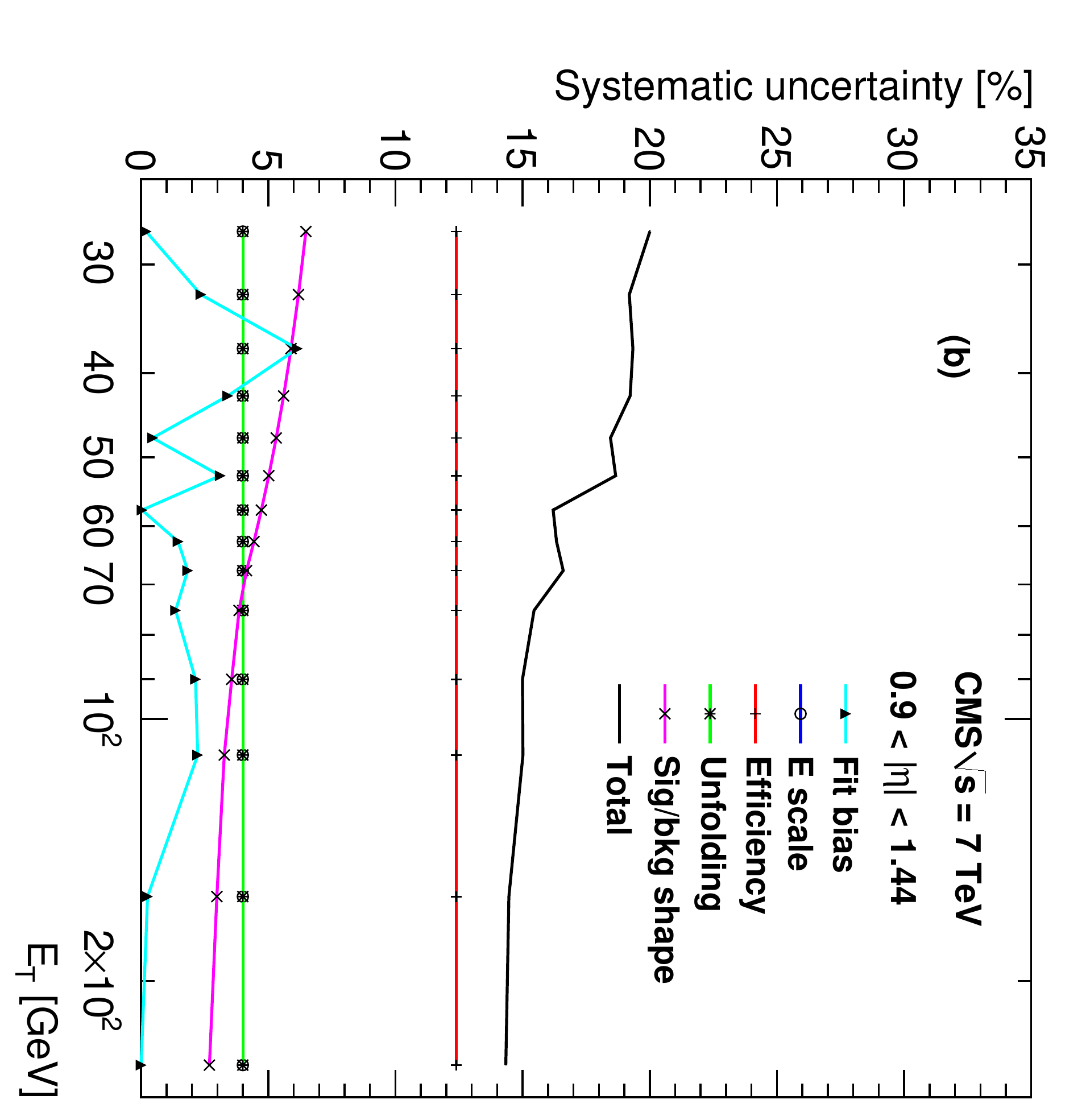} \\
\includegraphics[width=\figwid,angle=90]{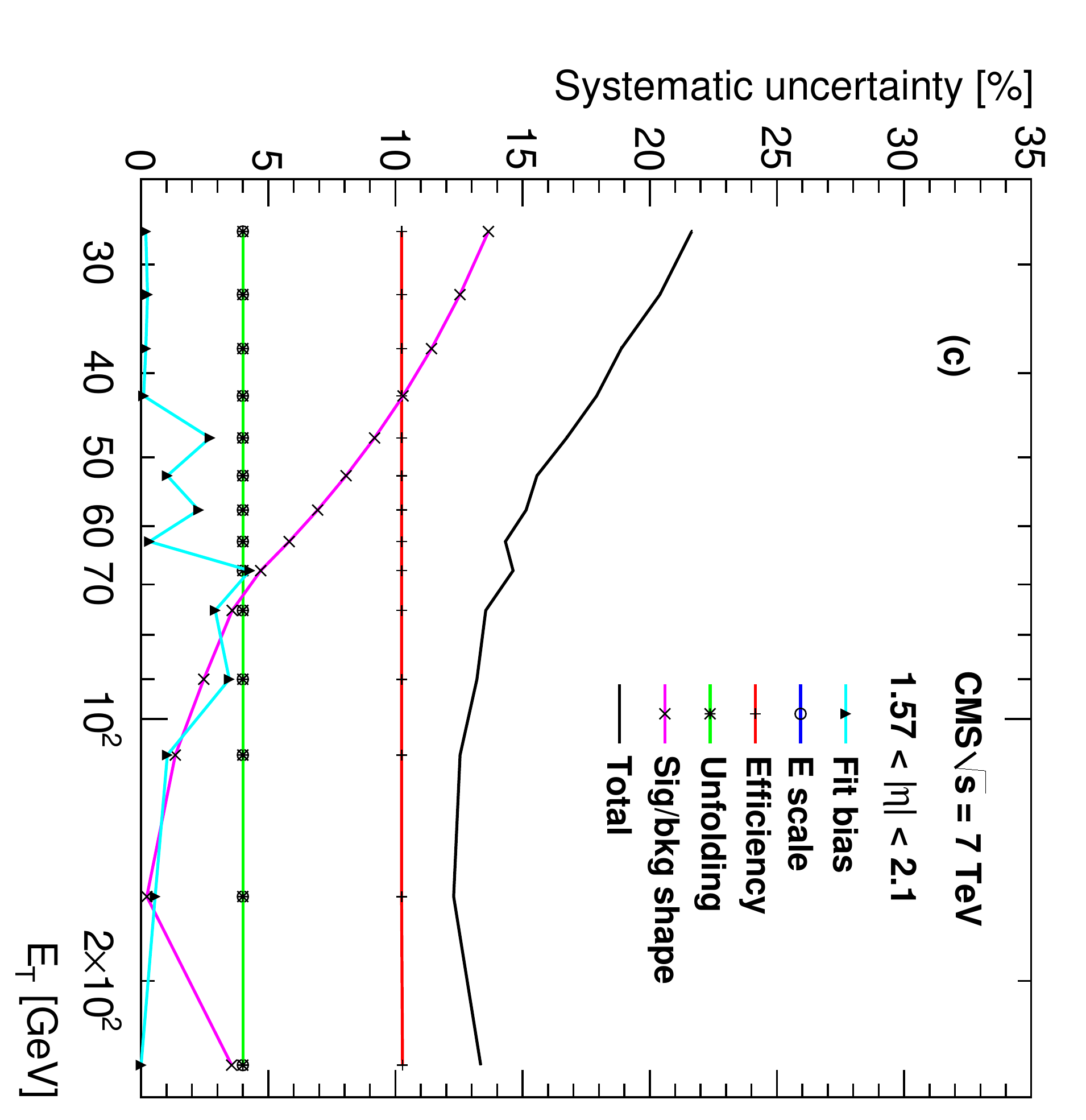}
\includegraphics[width=\figwid,angle=90]{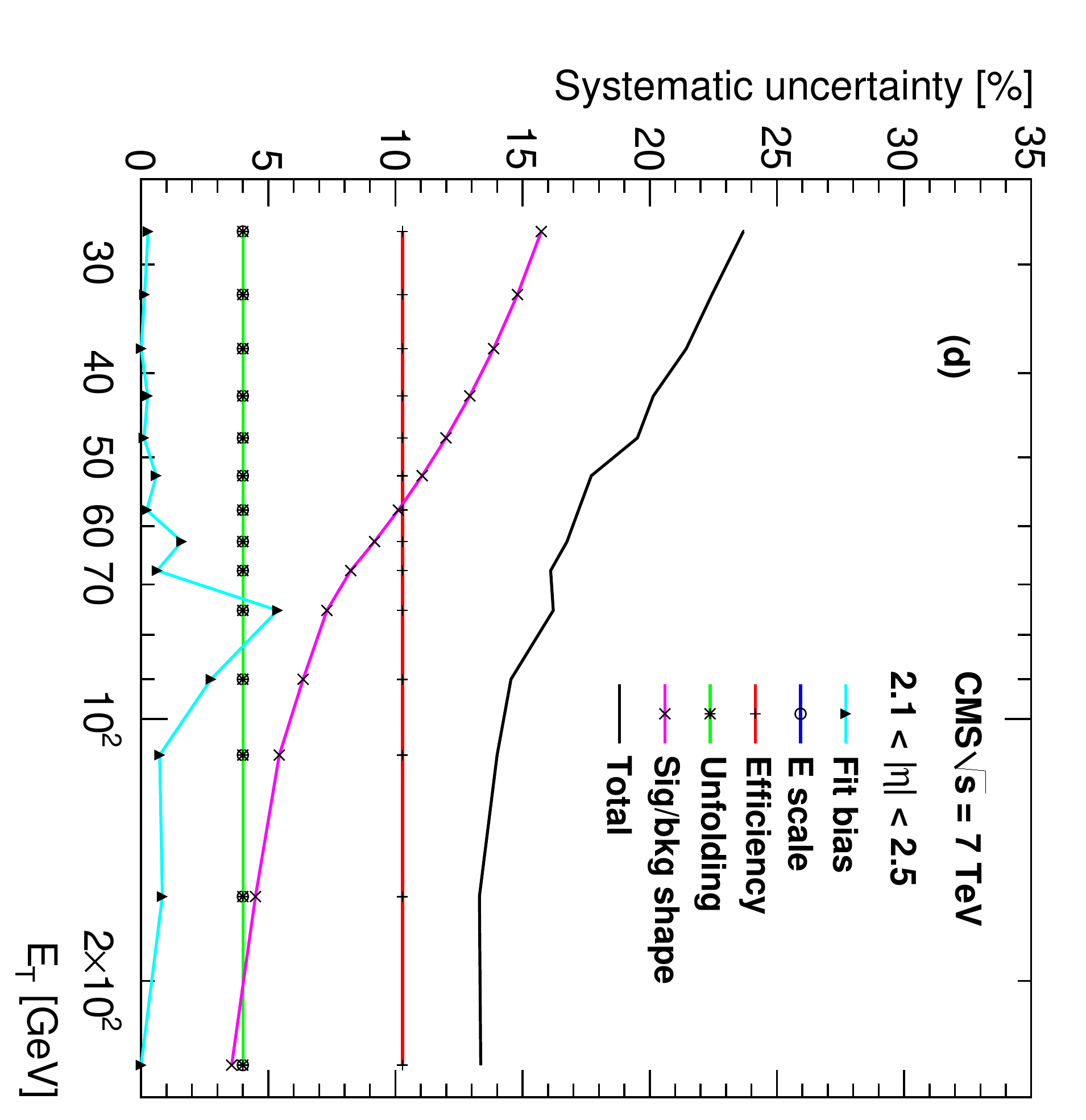}
\caption{Relative systematic uncertainties on the photon cross section
  measured with the photon conversion method in the
 four $\etagamma$ regions. Systematic uncertainties due to the uncertainties
 on the fit bias, energy scale, selection efficiency, unfolding correction
 factors, and signal and background shapes
 are shown, as well as their total quadrature sum (upper curve).
 }\label{fig:convSys}
\end{figure*}

Figure~\ref{fig:convSys} shows the $\etgamma$ dependence of each dominant
systematic uncertainty listed in Table~\ref{tab:final_sys} for the photon
cross section measured with the photon conversion method. The systematic
uncertainty associated with signal/background shape in the $\etagamma$ region
$|\etagamma| < 0.9$ increases with photon $\etgamma$, which is different
from the other $\etagamma$ regions for the following reasons. The size of the
systematic uncertainty is mainly driven by the difference between the simulated
 and observed background distributions with sideband
 selections; it is a balance between the decreasing number of background
events at high $\etgamma$ for this comparison (tending to make the uncertainty
larger) and increasing purity (tending to make the uncertainty lower).
In this $\etagamma$ bin the first effect dominates.

\subsection{Isolation method \label{sec:sysiso} }

For the highest-\etgamma\ bin in the isolation method (300--400\GeV), the
relative systematic uncertainty on $N^{\cPgg}$ is obtained from the
difference between the fitted purity at 200--300\GeV and the assumed 100\%
purity.
For the other \etgamma\ bins, the uncertainties associated with the signal and
background shapes arise from the uncertainties on the constrained values of
the shape parameters $a$, $p_0$, and $p_3$ in the likelihood fit
[Eqs.~(\ref{eq:sigPDF}) and (\ref{eq:bkgPDF}) in Section~\ref{sec:combIso}].

The constrained value of the signal parameter $a$ is varied by
$\pm30\%$ to account for the imperfect modelling of pile-up events; the value
30\% is the largest data-simulation difference observed among the four
$\eta$ bins in the electron sample. In order to estimate the effect due to the
modelling of nondirect photons in \PYTHIA, 
$2\rightarrow 2$ QCD processes
that contain ISR, FSR, and parton shower photons are removed in the signal
simulation, which results in a 5\% change in the constrained value of
parameter $a$.

The uncertainty on the background shape parameters $p_0$ and $p_3$ is mainly
driven by the size of background-enriched samples that are selected within the
sideband region of $\sigma_{\eta\eta}$ and that are used to derive
the data-to-simulation scaling factors. Because of the large statistical
uncertainties on the data-to-simulation scaling factors for higher photon
$\etgamma$ bins, the difference between the constrained values obtained by
applying and not applying the scaling factors is conservatively included
as a systematic uncertainty.

The uncertainty on $N^{\cPgg}$ from the signal and background shapes
is determined from pseudo-experiments by varying simultaneously the values
of parameters $a$, $p_0$, and $p_3$ according to their uncertainties.

\begin{figure*}[htbpH]
\centering
\includegraphics[width=\figwid,angle=90]{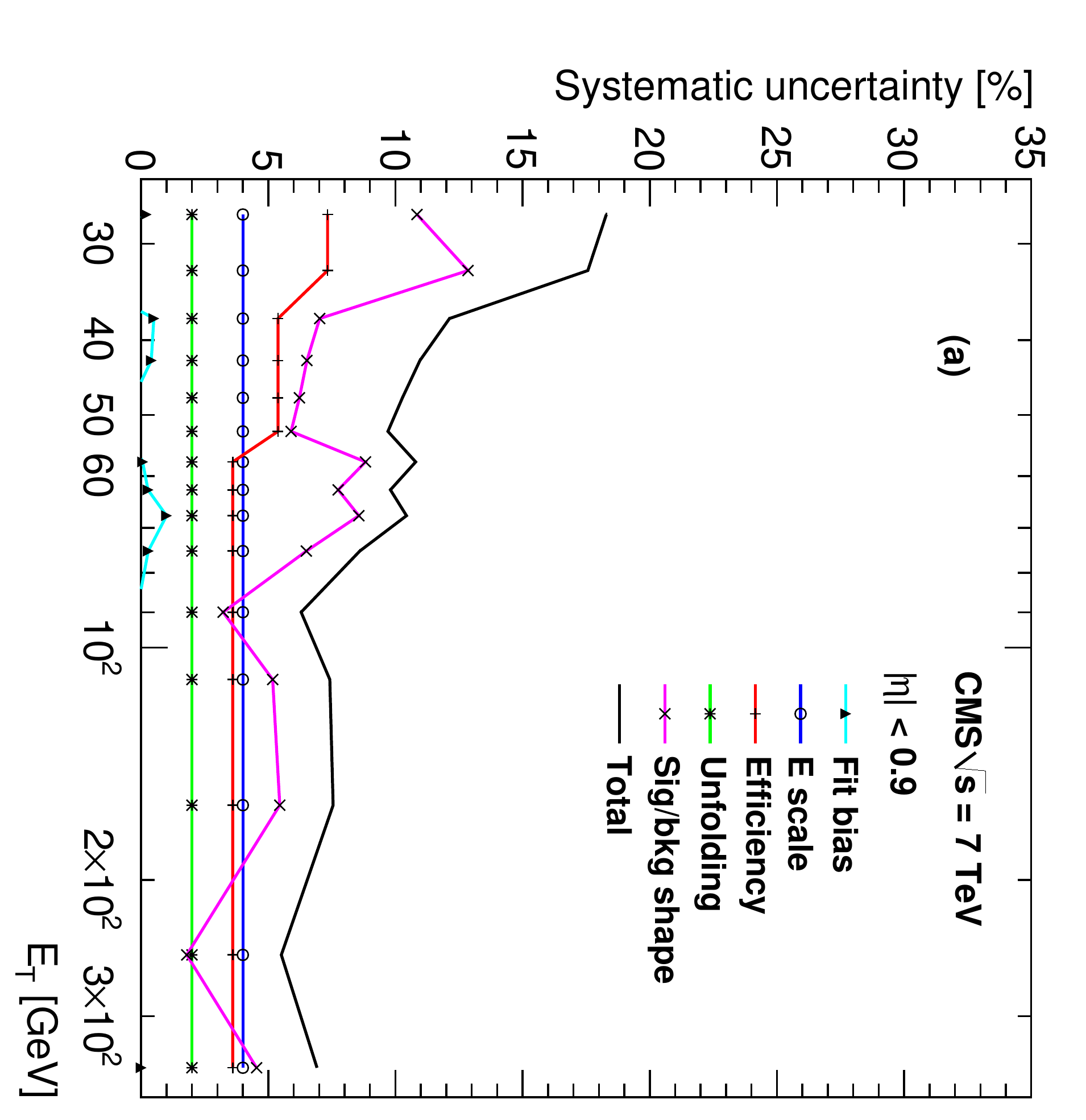}
\includegraphics[width=\figwid,angle=90]{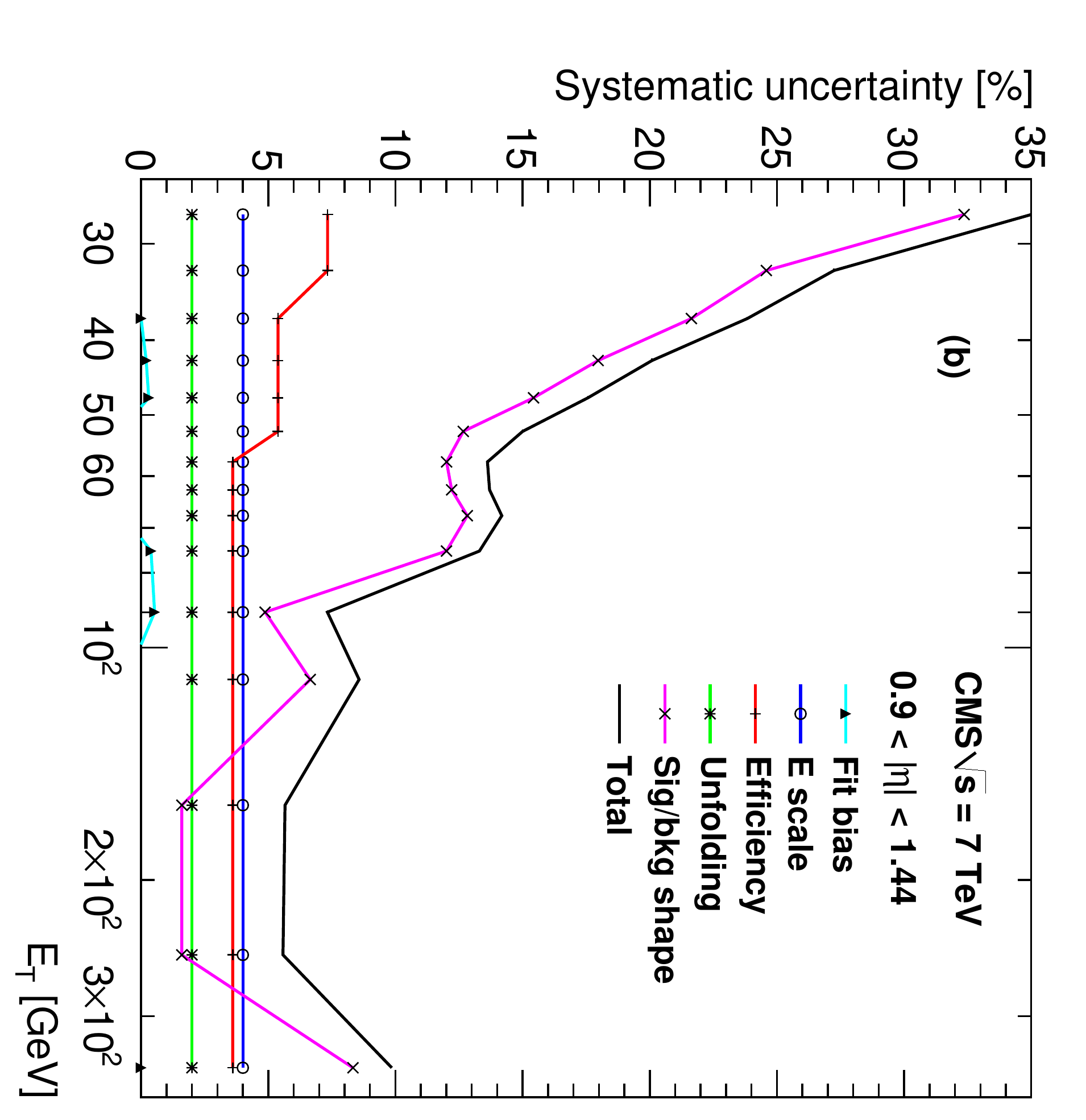}\\
\includegraphics[width=\figwid,angle=90]{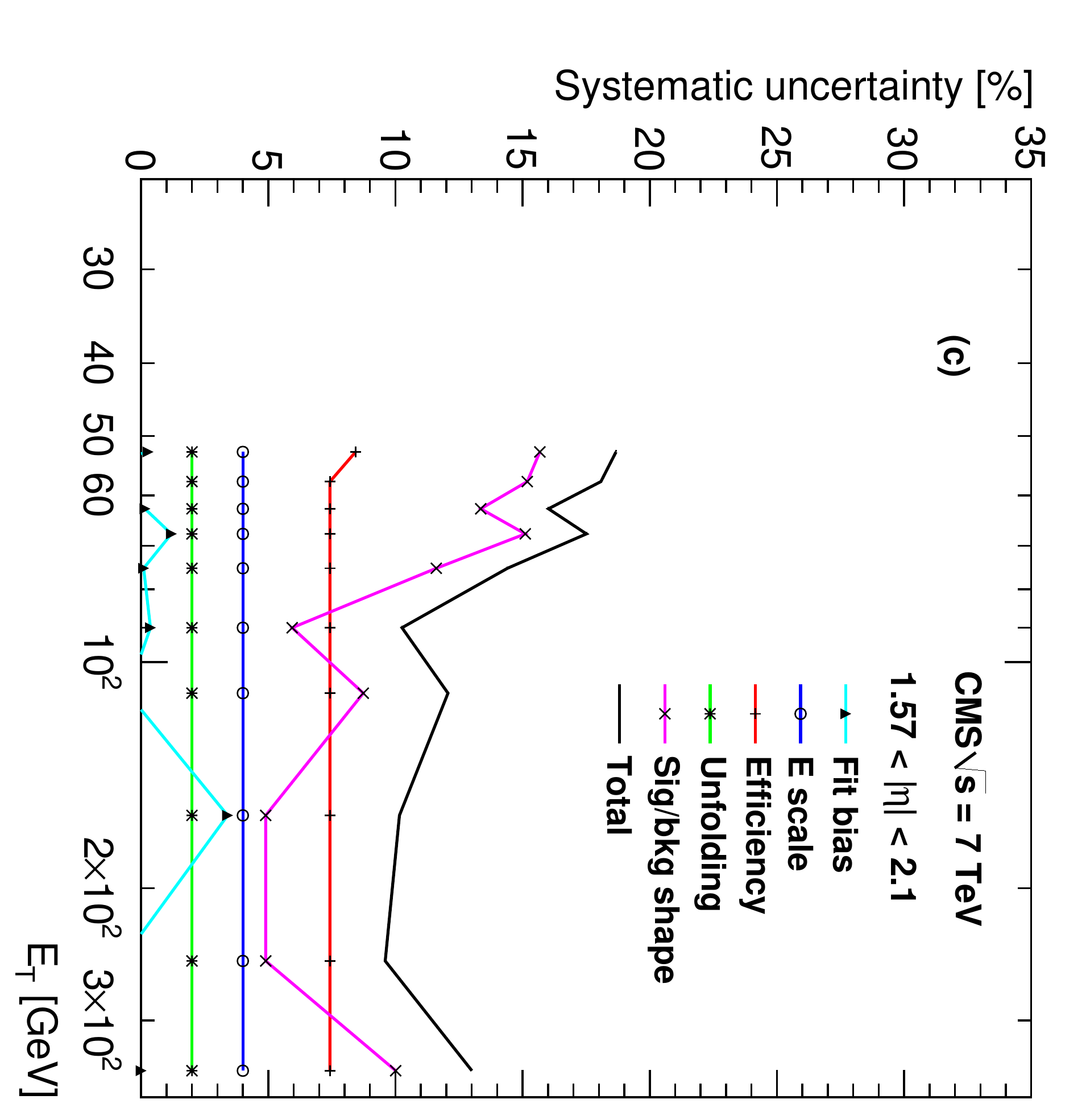}
\includegraphics[width=\figwid,angle=90]{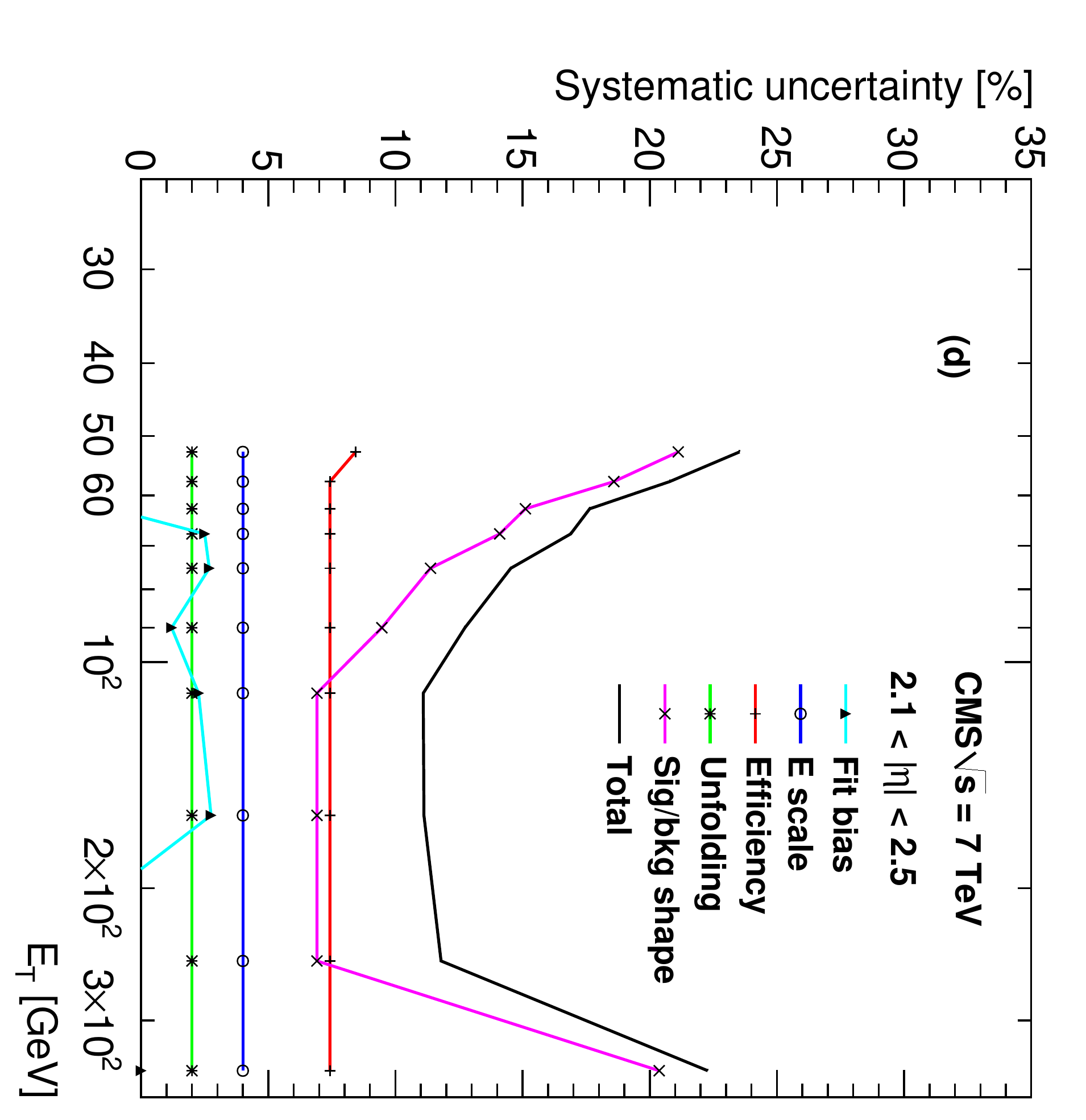}
\caption{Relative systematic uncertainties on the photon cross section
  measured with the isolation method in the four $\etagamma$ regions.
  Systematic uncertainties due to the uncertainties on the
 fit bias, energy scale, selection efficiency, unfolding correction factors,
 and signal and background shapes
 are shown, as well as their total quadrature sum (upper curve).
 }\label{fig:isoSys}
\end{figure*}

Figure~\ref{fig:isoSys} shows the $\etgamma$ dependence of each dominant
systematic uncertainty listed in Table~\ref{tab:final_sys} for the photon
cross section measured with the isolation method.
Variations of the systematic uncertainty associated with the
signal/background shapes are observed for the following reasons. In general,
the uncertainty decreases
with photon $\etgamma$, which is expected due to the increase of photon
purity. However, the difference
between simulated and observed $\Iso$ distribution increases with
$\etgamma$. For $\etgamma<55\GeV$, only data before the improvement of
LHC instantaneous luminosity are used (Table~\ref{tab:lumi}) and
in this period there are fewer pile-up collisions on average. Therefore, a
step or a transition in systematic uncertainty is observed at $\etgamma$ =
55\GeV in Figs.~\ref{fig:isoSys}~(a) and (b).
In addition, due to the lack of high-$\etgamma$ photon candidates that satisfy
sideband selection criteria, a background-enriched sample with
photon $\etgamma$ = 80--100\GeV is used to derive the data-to-simulation
scaling factors for all the $\etgamma$ bins above 80\GeV. Therefore, for
three out of four $\etagamma$ regions, a discontinuity in the
systematic uncertainty is observed at 100\GeV.
The systematic uncertainty on the cross section for low-$\etgamma$ photons in
the outer barrel is larger than in the central barrel because of a larger
difference between the simulated and observed isolation distributions.

\section{Results \label{sec:results}}

The differential cross section is defined as
\begin{equation}
d^{2}\sigma/d\etgamma d\etagamma = N^\cPgg \cdot \mathcal{U}/
(L\cdot\epsilon\cdot\Delta\etgamma\cdot\Delta\etagamma),
\label{eq:diffxsec}
\end{equation}
where $N^\cPgg$ is the signal photon yield measured from data
(Section~\ref{sec:yield}),
$L$ is the integrated luminosity, $\mathcal{U}$ denotes the bin-by-bin
unfolding correction factors, $\epsilon$ is the product of the efficiencies
(Section~\ref{sec:eff}), and $\Delta\etgamma$ and $\Delta\etagamma$ are the
sizes of the $\etgamma$ and $\etagamma$ bins.

The $\et$ of a photon candidate can be mismeasured because of
detector resolution and imperfections in the reconstruction algorithm.
The bin-by-bin unfolding
correction $\mathcal{U}$ is applied to account for these effects.
The correction is obtained from simulation for each $\etgamma$-$\etagamma$
bin, by taking the ratio of the generator- to the reconstruction-level
photon $\etgamma$ spectrum. Direct photons, simulated by \PYTHIA as described
in Section~\ref{sec:sample}, are used to derive the correction.
The difference in the correction factors obtained by using the $\etgamma$
spectrum of direct photons in \PYTHIA and by using that of the NLO pQCD
predictions~\cite{jetphox,jetphox2} is taken as a systematic uncertainty.
The resulting relative change on the cross section is listed in
Table~\ref{tab:final_sys} and shown in Figs.~\ref{fig:convSys} and
\ref{fig:isoSys} for the photon conversion and isolation methods, respectively.

After unfolding corrections are applied, the results of the photon conversion
and isolation methods are compared for the $\etgamma$ bins where both
are available.
The consistency of the results are quantified by the
global $\chi^2$ divided by the number of $\etgamma$ bins; they are
11.8/13, 20.4/13, 12.2/8, and 3.6/8 for the four pseudorapidity intervals,
respectively,
which indicates a good agreement between the results of the two methods.
The results of the two methods are combined using the procedure
described in Ref.~\cite{Lyons:1988rp}, weighting each method by its
corresponding uncertainty in each $\etgamma$-$\etagamma$ bin.
The weights are obtained by inverting the covariance matrix between the two
methods, which has elements $C_{ij} = \rho_{ij}\sigma_{i}\sigma_{j}$,
where $\sigma_{i}$ and $\sigma_{j}$ are the total uncertainties for the two
methods and $\rho_{ij}$ is the correlation coefficient.

Since the value of $\rho_{ij}$ is not known in data, the following
procedure is adopted for combining the results.
The systematic uncertainties from trigger efficiency and energy scale are
fully correlated  because the same procedure is applied to both
measurements and they give a lower limit on $\rho_{ij}$.
The systematic uncertainties due to signal and background
shapes as well as to the fitting bias
have negligible correlation, as checked from simulation studies; they
provide the upper limit on the correlation coefficients $\rho_{ij}$.
For the remaining sources of uncertainties, the correlation coefficients are
varied from zero to one. Each unknown coefficient is varied independently,
and for each variation, the resulting value of $\rho_{ij}$ is used in the
combination. The final value of the combined cross section is the mean
of the central values obtained for each $\rho_{ij}$.
For each value of $\rho_{ij}$ the uncertainty on the combined measurement is
evaluated and the final, total uncertainty is conservatively quoted as the
mean of the resulting uncertainty distribution plus its standard deviation.

Since the major sources of systematic uncertainties  (signal and
background shapes) are uncorrelated, the impact on the final combined result
of varying the correlations on the remaining sources is limited to the percent
level.
The central values and uncertainties generated with this procedure were
studied in pseudo-experiments to confirm that they have the desired
properties.

Since the conversion method has smaller uncertainty than the isolation method
at low \etgamma, it receives a higher weight in the combination at low
\etgamma. The situation is reversed at high \etgamma.
Table~\ref{tab:xs} lists the final measured cross section with corresponding
statistical and systematic uncertainties.

\begin{table}[htbpH]\renewcommand{\arraystretch}{1.3}
  \begin{center}
    \caption{
    Measured isolated prompt photon differential cross section
    $d^{2}\sigma/d\etgamma d\etagamma$ in the four pseudorapidity regions.
    The quoted uncertainties are statistical and systematic, respectively.
   \label{tab:xs} }
  \begin{tabular}{ l  r  r} \hline \hline
   & $|\etagamma|<0.9$ & $0.9<|\etagamma|<1.44$\\\hline
   $\etgamma$ (\!\GeV)    &  Cross section (nb/\!\GeV)  & Cross section (nb/\!\GeV)\\ \hline

 25--30     & $(7.83 \pm 0.17^{+0.96}_{-0.96}) \times 10^{-1}$ & $(6.69 \pm 0.85^{+1.25}_{-1.33}) \times 10^{-1}$\\
 30--35     & $(3.85 \pm 0.14^{+0.46}_{-0.42}) \times 10^{-1}$ & $(4.07 \pm 0.46^{+0.67}_{-0.55}) \times 10^{-1}$\\
 35--40     & $(2.04 \pm 0.04^{+0.19}_{-0.19}) \times 10^{-1}$ & $(1.90 \pm 0.19^{+0.31}_{-0.29}) \times 10^{-1}$\\
 40--45     & $(1.25 \pm 0.03^{+0.11}_{-0.11}) \times 10^{-1}$ & $(1.42 \pm 0.06^{+0.19}_{-0.19}) \times 10^{-1}$\\
 45--50     & $(7.93 \pm 0.22^{+0.63}_{-0.66}) \times 10^{-2}$ & $(7.81 \pm 0.48^{+0.96}_{-0.93}) \times 10^{-2}$\\
 50--55     & $(4.97 \pm 0.16^{+0.37}_{-0.39}) \times 10^{-2}$ & $(5.03 \pm 0.24^{+0.54}_{-0.54}) \times 10^{-2}$\\
 55--60     & $(3.49 \pm 0.09^{+0.28}_{-0.28}) \times 10^{-2}$ & $(3.46 \pm 0.16^{+0.31}_{-0.34}) \times 10^{-2}$\\
 60--65     & $(2.31 \pm 0.09^{+0.17}_{-0.18}) \times 10^{-2}$ & $(2.18 \pm 0.12^{+0.18}_{-0.22}) \times 10^{-2}$\\
 65--70     & $(1.61 \pm 0.07^{+0.12}_{-0.13}) \times 10^{-2}$ & $(1.58 \pm 0.10^{+0.14}_{-0.15}) \times 10^{-2}$\\
 70--80     & $(1.05 \pm 0.03^{+0.09}_{-0.08}) \times 10^{-2}$ & $(1.09 \pm 0.05^{+0.10}_{-0.10}) \times 10^{-2}$\\
 80--100    & $(4.80 \pm 0.12^{+0.30}_{-0.28}) \times 10^{-3}$ & $(4.86 \pm 0.24^{+0.37}_{-0.35}) \times 10^{-3}$\\
 100--120   & $(1.89 \pm 0.05^{+0.14}_{-0.14}) \times 10^{-3}$ & $(2.00 \pm 0.10^{+0.21}_{-0.17}) \times 10^{-3}$\\
 120--200   & $(4.07 \pm 0.16^{+0.33}_{-0.29}) \times 10^{-4}$ & $(4.06 \pm 0.16^{+0.23}_{-0.23}) \times 10^{-4}$\\
 200--300   & $(4.00 \pm 0.29^{+0.27}_{-0.27}) \times 10^{-5}$ & $(3.60 \pm 0.39^{+0.20}_{-0.20}) \times 10^{-5}$\\
 300--400   & $(8.20 \pm 1.22^{+0.59}_{-0.54}) \times 10^{-6}$ & $(7.84 \pm 1.53^{+0.75}_{-0.75}) \times 10^{-6}$\\

  \hline\hline
   & $1.57<|\etagamma|<2.1$ & $2.1<|\etagamma|<2.5$\\\hline
   $\etgamma$ (\!\GeV)    &  Cross section (nb/\!\GeV) & Cross section (nb/\!\GeV) \\ \hline

 25--30     & $(6.39 \pm 0.50^{+1.35}_{-1.35}) \times 10^{-1}$ & $(7.40 \pm 0.58^{+1.74}_{-1.70}) \times 10^{-1}$\\
 30--35     & $(2.93 \pm 0.34^{+0.59}_{-0.57}) \times 10^{-1}$ & $(4.09 \pm 0.38^{+0.91}_{-0.91}) \times 10^{-1}$\\
 35--40     & $(1.92 \pm 0.13^{+0.36}_{-0.36}) \times 10^{-1}$ & $(1.97 \pm 0.15^{+0.42}_{-0.41}) \times 10^{-1}$\\
 40--45     & $(1.18 \pm 0.10^{+0.20}_{-0.20}) \times 10^{-1}$ & $(1.22 \pm 0.12^{+0.24}_{-0.24}) \times 10^{-1}$\\
 45--50     & $(7.03 \pm 0.69^{+1.13}_{-1.13}) \times 10^{-2}$ & $(8.13 \pm 0.94^{+1.54}_{-1.54}) \times 10^{-2}$\\
 50--55     & $(5.63 \pm 0.31^{+0.63}_{-0.58}) \times 10^{-2}$ & $(5.34 \pm 0.42^{+0.70}_{-0.65}) \times 10^{-2}$\\
 55--60     & $(3.72 \pm 0.21^{+0.42}_{-0.34}) \times 10^{-2}$ & $(3.10 \pm 0.22^{+0.39}_{-0.34}) \times 10^{-2}$\\
 60--65     & $(2.34 \pm 0.14^{+0.24}_{-0.18}) \times 10^{-2}$ & $(2.39 \pm 0.12^{+0.27}_{-0.27}) \times 10^{-2}$\\
 65--70     & $(1.61 \pm 0.12^{+0.17}_{-0.13}) \times 10^{-2}$ & $(1.75 \pm 0.09^{+0.19}_{-0.18}) \times 10^{-2}$\\
 70--80     & $(1.07 \pm 0.06^{+0.10}_{-0.09}) \times 10^{-2}$ & $(1.03 \pm 0.05^{+0.11}_{-0.11}) \times 10^{-2}$\\
80--100     & $(4.91 \pm 0.14^{+0.35}_{-0.37}) \times 10^{-3}$ & $(3.86 \pm 0.18^{+0.33}_{-0.33}) \times 10^{-3}$\\
100--120    & $(1.48 \pm 0.12^{+0.13}_{-0.07}) \times 10^{-3}$ & $(1.39 \pm 0.08^{+0.12}_{-0.12}) \times 10^{-3}$\\
120--200    & $(3.68 \pm 0.16^{+0.27}_{-0.27}) \times 10^{-4}$ & $(2.29 \pm 0.19^{+0.18}_{-0.18}) \times 10^{-4}$\\
200--300    & $(2.80 \pm 0.39^{+0.30}_{-0.30}) \times 10^{-5}$ & $(1.40 \pm 0.30^{+0.26}_{-0.26}) \times 10^{-5}$\\
300--400    & $(2.80 \pm 0.99^{+0.37}_{-0.34}) \times 10^{-6}$ & $(5.42 \pm 5.42^{+1.21}_{-1.21}) \times 10^{-7}$ \\

  \hline\hline
  \end{tabular}
\end{center}
\end{table}

\section{Comparison with Theory \label{sec:dataToTheory}}

\begin{figure}[hbtp]
\centering

\includegraphics[width=\onecolfigwid]{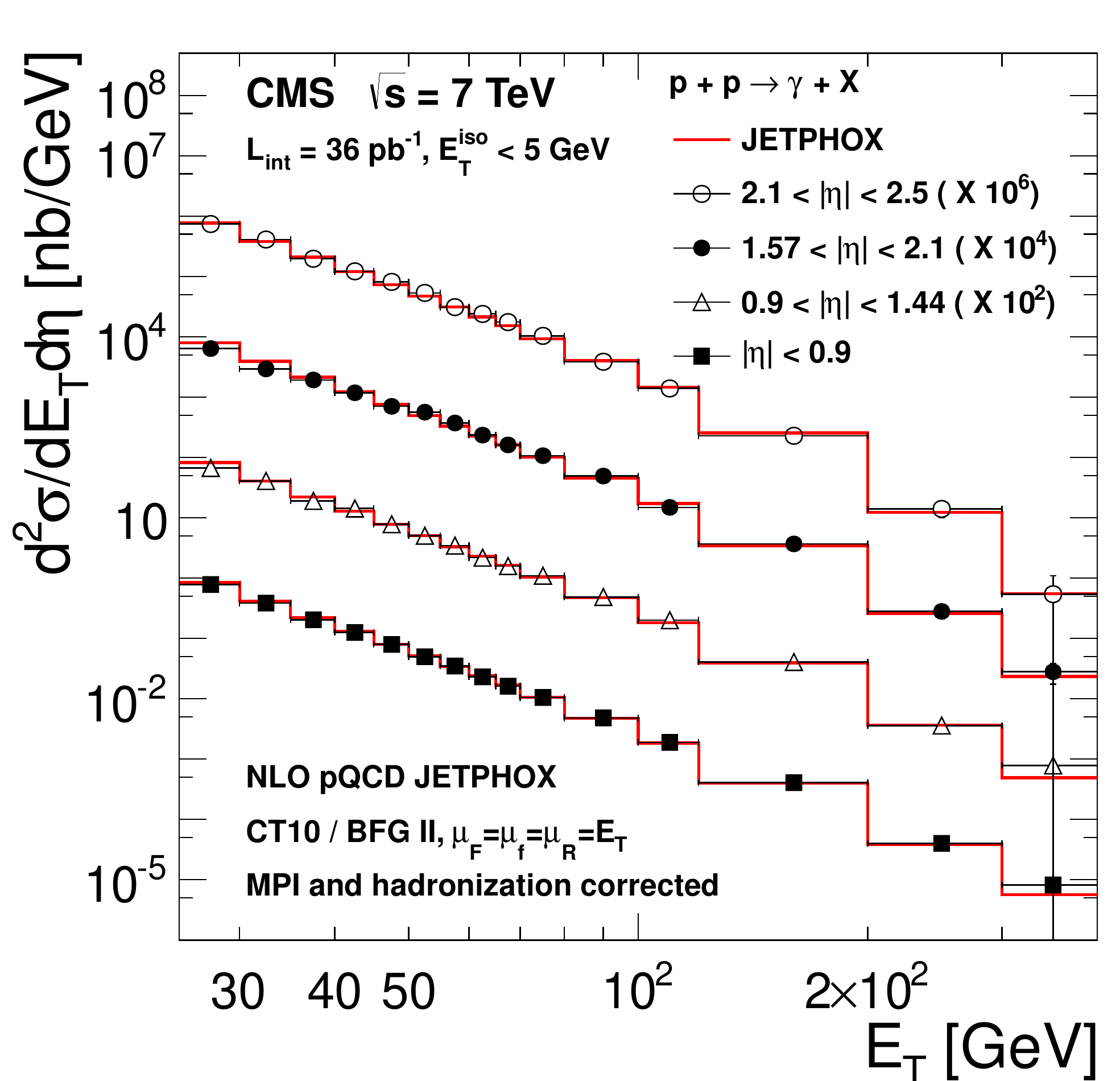}
\caption{Measured isolated prompt photon differential cross sections
         (markers) as a function of transverse energy in the four
	 pseudorapidity regions and the predictions from \jetphox~1.3.0
         using the CT10 PDFs (histograms). The error bars are the
	 quadrature sums of statistical and systematic uncertainties
	 on the measurements. The cross
	 sections are scaled by the factors shown in the legend for easier
	 viewing.}
\label{fig:data}
\end{figure}

The measured differential cross sections are shown in Fig.~\ref{fig:data}
for the four pseudorapidity ranges considered,
together with NLO pQCD predictions from \jetphox~1.3.0~\cite{jetphox,jetphox2}
using the CT10 PDFs~\cite{ct10} and the BFG
set II of fragmentation functions (FFs)~\cite{BFG}.
The $4\%$ overall uncertainty on the integrated luminosity is considered
separately.
The hadronic energy surrounding the photon is required to be
at most $5\GeV$ within $R<0.4$ at the parton level.
The renormalisation, factorisation, and fragmentation scales ($\mu_{R}$,
$\mu_{F}$, and $\mu_{f}$) are all set to the $\etgamma$ of photon.
To estimate the effect of the choice of theory scales in the predictions, the
three scales are varied independently between $\etgamma/2$ and $2\etgamma$
while keeping the ratio of one scale to the other scales, or vice versa, at
most two.
Retaining the largest cross-section variation at each $\etgamma$ bin, the
predictions change by $\pm22\%$ to $\pm7\%$ with increasing $\etgamma$.
The uncertainty on the predictions due to the PDFs is determined from
the 52+1 CT10 PDF sets using the Hessian method~\cite{Hessian,pdf4lhcreport}
with a reduction of a factor of 1.645 to obtain the 68\% confidence level
(CL) variation.
The uncertainty due to the variation of $\alpS(M_Z)$ values is estimated
from the difference between CT10 PDFs with $\alpS(M_Z)$ set to 0.118, and
two CT10as sets with $\alpS(M_Z)$ set to $0.118\pm0.001$
corresponding to the 68\% CL variation. The $\alpS(M_Z)$ uncertainty is
added in quadrature with the PDF uncertainty~\cite{ct10};
the combined PDF and $\alpS$ uncertainties are within the ranges of
2.5--8.0\%, 1.6--8.2\%, 2.4--8.5\%, and 1.7--11\% in the four $\etagamma$
regions, respectively.
Finally, using the BFG set I of FFs instead of the BFG set II yields
negligible differences in the predictions.

The theoretical predictions are multiplied by an additional correction
factor $C$ to account for the presence of contributions from the
underlying event and parton-to-hadron fragmentation, which tend to
increase the hadronic energy inside the isolation cone.
Using simulated \PYTHIA events, $C$ is determined as
the ratio between the isolated fraction of the total prompt photon cross
section at the hadron level and the same fraction obtained after turning off
both multiple-parton interactions (MPI) and hadronization.
Four different sets of \PYTHIA parameters (Z2~\cite{pythia-z2},
D6T, DWT, and Perugia-0~\cite{pythia-p0})
are considered.
The average of $C$ over all parameter sets, $\bar{C}=0.975\pm 0.006$, has
little $\etgamma$ and $\etagamma$ dependence.
The uncertainty on $\bar{C}$ is the root mean square of the
results obtained with the different \PYTHIA parameter sets.
The correction reduces the predicted cross section, since the
presence of extra activity results in some photons failing the
isolation requirements.

\begin{figure*}[hbtp]
\centering
\includegraphics[width=\figwid]{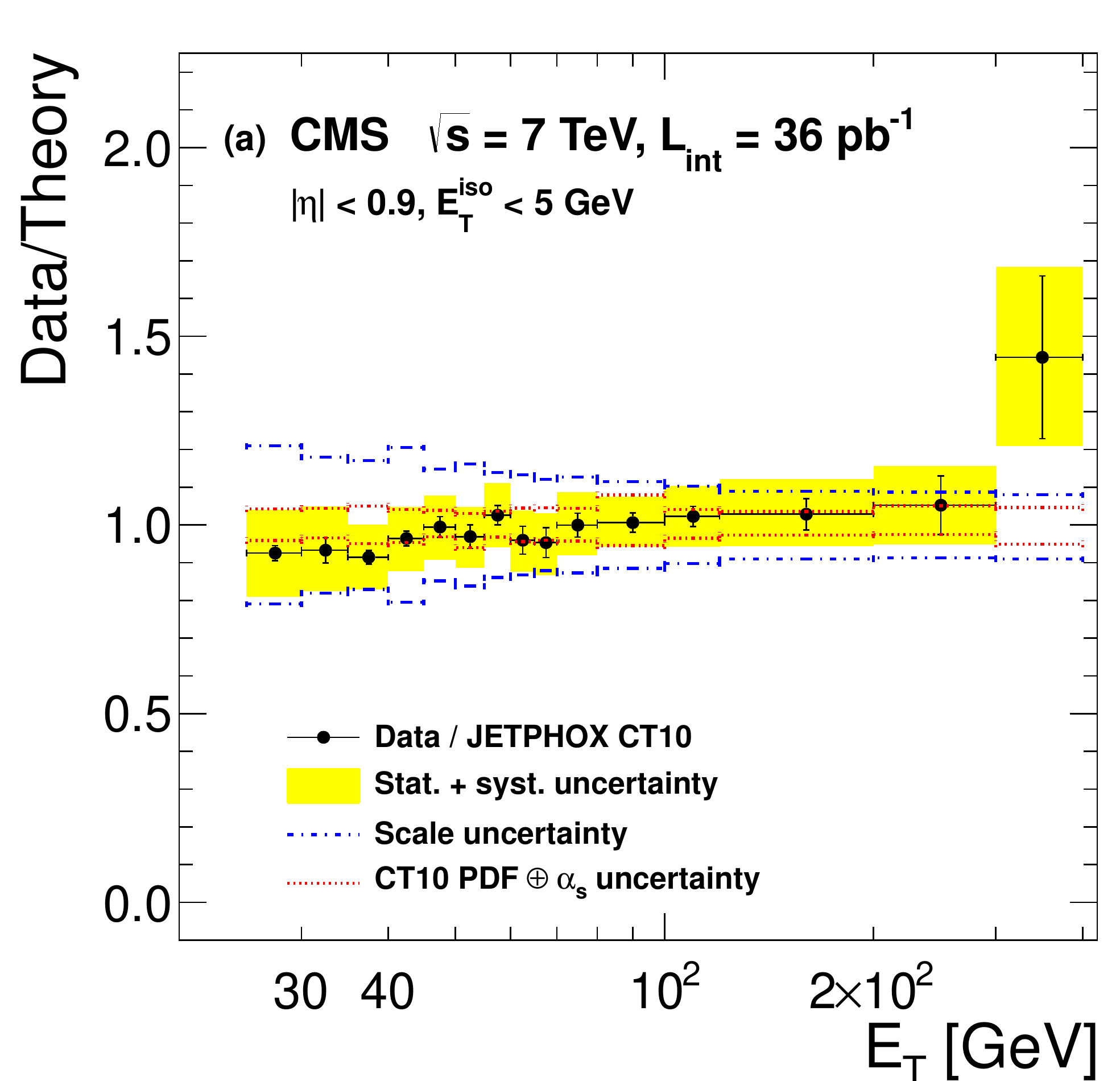}
\includegraphics[width=\figwid]{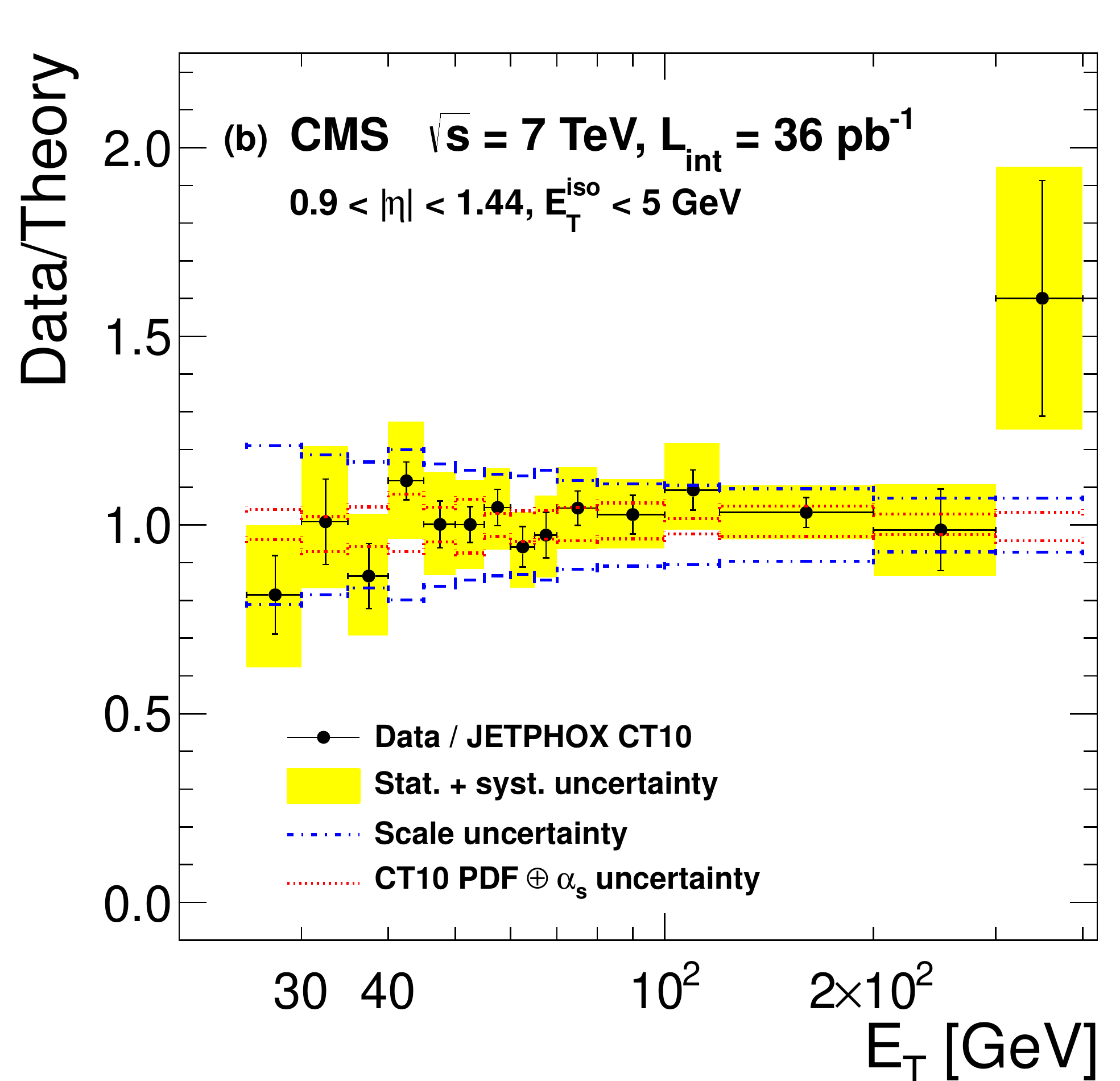}\\
\includegraphics[width=\figwid]{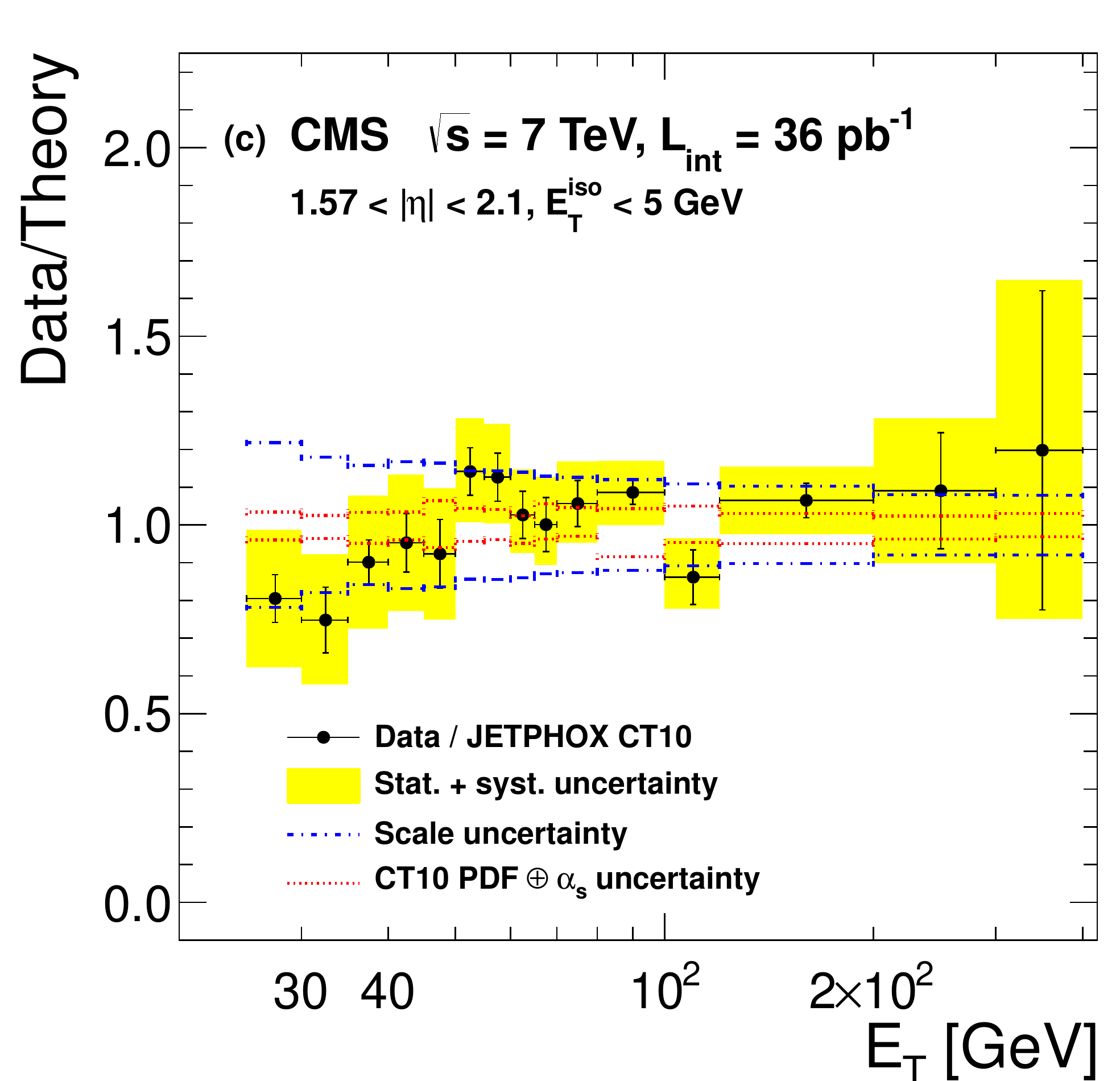}
\includegraphics[width=\figwid]{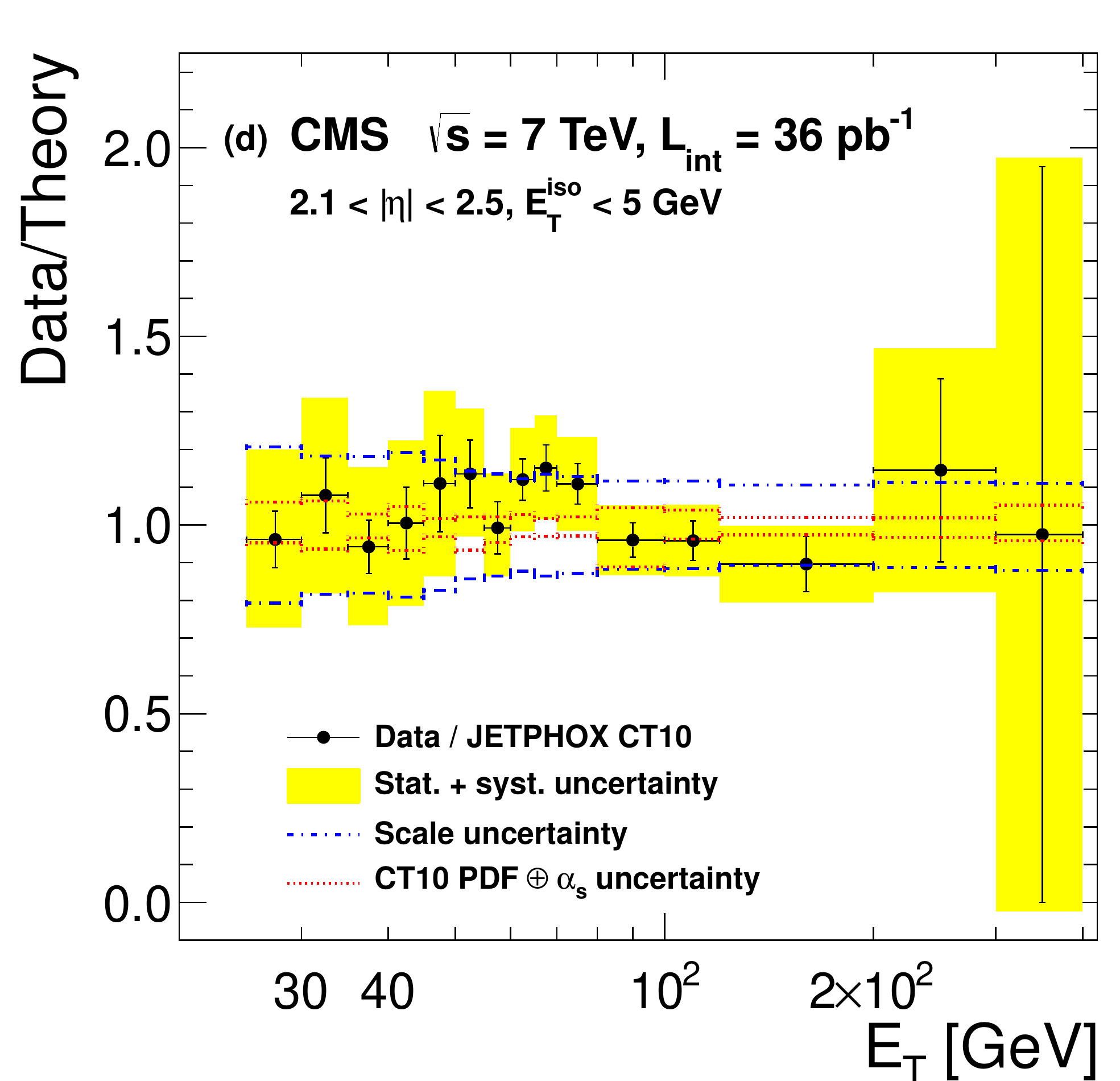}
\caption{
        Ratios of the measured isolated prompt photon differential cross
	section to the NLO pQCD predictions from \jetphox~1.3.0 using the CT10
	PDFs.
        The vertical error bars show the statistical uncertainties, while the
        shaded areas show the statistical and systematic
        uncertainties added in quadrature.
        The $4\%$ luminosity uncertainty on the data is not included. The two
        sets of curves show the uncertainties on the theoretical predictions
        due to their dependence on the renormalisation, factorisation, and
        fragmentation scales, and on the variation of CT10 $\alpS$ and PDFs.
        A correction to account for extra activity ($\bar{C}=0.975\pm 0.006$)
	is applied to the theoretical predictions, as explained in the text.
	}
\label{fig:data_theory}
\end{figure*}

Overall, predictions from the NLO pQCD calculations agree with
the measured cross sections within uncertainties, as shown in
Figs.~\ref{fig:data} and~\ref{fig:data_theory}. However,
for photons with lower $\etgamma$ in the $\etagamma$ regions
$|\etagamma|<0.9$, $0.9 < |\etagamma|< 1.44$, and $1.57 < |\etagamma|< 2.1$,
the cross sections predicted by NLO pQCD tend to be larger than the measured
cross sections, similar to the observation in Ref.~\cite{ref:ATLASInclPho}.

\section{Conclusion \label{sec:conclusion}}

A measurement of the differential cross section for the production of
isolated prompt photons with $\etgamma$ = 25--400\GeV in $\Pp\Pp$ collisions
at $\sqrt{s}=7\TeV$ has been performed in four intervals of
 pseudorapidity: $|\etagamma|<0.9$, $0.9 < |\etagamma|< 1.44$,
$1.57 < |\etagamma|< 2.1$, and $2.1 < |\etagamma|< 2.5$.
Two variables are explored to estimate the prompt photon yield: the ratio of
the energy measured in the electromagnetic calorimeter to the momentum measured
in the tracker for converted photons, and the isolation measured in the
tracker and calorimeters.
The differential cross sections obtained with these two methods are
combined into one measurement. Predictions from the NLO pQCD are found to
agree with the measured cross section within uncertainties,
although at low $\etgamma$ the predictions tend to be higher than the measured
cross section.
This measurement probes the kinematic region $0.007 < \xtgamma< 0.114$,
extends the previous CMS measurement to wider ranges of photon $\etgamma$ and
pseudorapidity, establishes a benchmark for photon identification and
background estimation, and determines the rate of one of the background
processes affecting searches for new physics involving photons.

\section*{Acknowledgement}

\hyphenation{Bundes-ministerium Forschungs-gemeinschaft Forschungs-zentren} We wish to congratulate our colleagues in the CERN accelerator departments for the excellent performance of the LHC machine. We thank the technical and administrative staff at CERN and other CMS institutes. This work was supported by the Austrian Federal Ministry of Science and Research; the Belgium Fonds de la Recherche Scientifique, and Fonds voor Wetenschappelijk Onderzoek; the Brazilian Funding Agencies (CNPq, CAPES, FAPERJ, and FAPESP); the Bulgarian Ministry of Education and Science; CERN; the Chinese Academy of Sciences, Ministry of Science and Technology, and National Natural Science Foundation of China; the Colombian Funding Agency (COLCIENCIAS); the Croatian Ministry of Science, Education and Sport; the Research Promotion Foundation, Cyprus; the Estonian Academy of Sciences and NICPB; the Academy of Finland, Finnish Ministry of Education and Culture, and Helsinki Institute of Physics; the Institut National de Physique Nucl\'eaire et de Physique des Particules~/~CNRS, and Commissariat \`a l'\'Energie Atomique et aux \'Energies Alternatives~/~CEA, France; the Bundesministerium f\"ur Bildung und Forschung, Deutsche Forschungsgemeinschaft, and Helmholtz-Gemeinschaft Deutscher Forschungszentren, Germany; the General Secretariat for Research and Technology, Greece; the National Scientific Research Foundation, and National Office for Research and Technology, Hungary; the Department of Atomic Energy and the Department of Science and Technology, India; the Institute for Studies in Theoretical Physics and Mathematics, Iran; the Science Foundation, Ireland; the Istituto Nazionale di Fisica Nucleare, Italy; the Korean Ministry of Education, Science and Technology and the World Class University program of NRF, Korea; the Lithuanian Academy of Sciences; the Mexican Funding Agencies (CINVESTAV, CONACYT, SEP, and UASLP-FAI); the Ministry of Science and Innovation, New Zealand; the Pakistan Atomic Energy Commission; the State Commission for Scientific Research, Poland; the Funda\c{c}\~ao para a Ci\^encia e a Tecnologia, Portugal; JINR (Armenia, Belarus, Georgia, Ukraine, Uzbekistan); the Ministry of Science and Technologies of the Russian Federation, the Russian Ministry of Atomic Energy and the Russian Foundation for Basic Research; the Ministry of Science and Technological Development of Serbia; the Ministerio de Ciencia e Innovaci\'on, and Programa Consolider-Ingenio 2010, Spain; the Swiss Funding Agencies (ETH Board, ETH Zurich, PSI, SNF, UniZH, Canton Zurich, and SER); the National Science Council, Taipei; the Scientific and Technical Research Council of Turkey, and Turkish Atomic Energy Authority; the Science and Technology Facilities Council, UK; the US Department of Energy, and the US National Science Foundation.

Individuals have received support from the Marie-Curie programme and the European Research Council (European Union); the Leventis Foundation; the A. P. Sloan Foundation; the Alexander von Humboldt Foundation; the Associazione per lo Sviluppo Scientifico e Tecnologico del Piemonte (Italy); the Belgian Federal Science Policy Office; the Fonds pour la Formation \`a la Recherche dans l'Industrie et dans l'Agriculture (FRIA-Belgium); the Agentschap voor Innovatie door Wetenschap en Technologie (IWT-Belgium); and the Council of Science and Industrial Research, India.
\clearpage
\bibliography{auto_generated}   

\cleardoublepage \appendix\section{The CMS Collaboration \label{app:collab}}\begin{sloppypar}\hyphenpenalty=5000\widowpenalty=500\clubpenalty=5000\input{QCD-10-037-authorlist.tex}\end{sloppypar}
\end{document}

%% file: QCD-10-037-authorlist.tex
\textbf{Yerevan Physics Institute,  Yerevan,  Armenia}\\*[0pt]
S.~Chatrchyan, V.~Khachatryan, A.M.~Sirunyan, A.~Tumasyan
\vskip\cmsinstskip
\textbf{Institut f\"{u}r Hochenergiephysik der OeAW,  Wien,  Austria}\\*[0pt]
W.~Adam, T.~Bergauer, M.~Dragicevic, J.~Er\"{o}, C.~Fabjan, M.~Friedl, R.~Fr\"{u}hwirth, V.M.~Ghete, J.~Hammer\cmsAuthorMark{1}, S.~H\"{a}nsel, M.~Hoch, N.~H\"{o}rmann, J.~Hrubec, M.~Jeitler, W.~Kiesenhofer, M.~Krammer, D.~Liko, I.~Mikulec, M.~Pernicka, B.~Rahbaran, H.~Rohringer, R.~Sch\"{o}fbeck, J.~Strauss, A.~Taurok, F.~Teischinger, C.~Trauner, P.~Wagner, W.~Waltenberger, G.~Walzel, E.~Widl, C.-E.~Wulz
% ...
\vskip\cmsinstskip
% ...
1:~~Also at CERN, European Organization for Nuclear Research, Geneva, Switzerland\\
% ...
